\documentclass{aa}
\usepackage{txfonts}
\usepackage{graphicx}
\usepackage{aalongtable}
\usepackage{lscape}
\usepackage{txfonts}
\usepackage{natbib}
\usepackage{color}
\newcommand{\sext}{\texttt{SExtractor}}
\newcommand{\bv}{$(B-V)$}
\newcommand{\fwhm}{\texttt{FWHM}}
\newcommand{\mumax}{\ensuremath{\mu_{max}}}
\newcommand{\fwhms}{\ensuremath{\texttt{FWHM}_*}}
\newcommand{\sbg}{\ensuremath{\sigma_{bg}}}

\defcitealias{Fasano2006}{Paper~I} 

\begin{document}
   \title{WINGS: A WIde-field Nearby Galaxy-cluster Survey. II}
   
   \subtitle{Deep optical  photometry  of  77  nearby  clusters
     \thanks{Based on observations taken at the Issac Newton Telescope
       (2.5m-INT) sited  at Roque de  los Muchachos (La Palma, Spain),
       and the MPG/ESO-2.2m Telescope sited at La Silla (Chile).}}

   \author{
 J. Varela\inst{1,2} \and
 M. D'Onofrio\inst{3} \and 
 C. Marmo \inst{4} \and
 G. Fasano\inst{1} \and 
 D. Bettoni\inst{1} \and 
 A. Cava\inst{1,5} \and 
 W.J. Couch\inst{6}\and 
 A. Dressler\inst{7} \and
 P. Kj{\ae}rgaard\inst{8} \and 
 M. Moles\inst{2} \and
 E. Pignatelli\inst{1} \and 
 B.M. Poggianti \inst{1} \and
 T. Valentinuzzi\inst{3}
          }

   \institute{
     INAF -- Padova Astronomical Observatory, Vicolo Osservatorio 5,
     35122 Padova, Italy \and
     Instituto de Astrof\'{\i}sica de Andaluc\'{\i}a (C.S.I.C.) Apartado
     3004, 18080 Granada, Spain \and
     Astronomy Department, University of Padova, Vicolo Osservatorio 2,
     35122 Padova, Italy \and
     UMR 8148 IDES Interactions et Dynamique des Environnements de Surface Universit\'e,
      Paris-Sud, 91405 ORSAY Cedex, France
        \and
     Instituto de Astrof\'{\i}sica de Canarias, c/ V\'{\i}a L\'actea
     s/n, La Laguna, Spain \and
     Centre for Astrophysics and Supercomputing, Swinburne University of Technology,
     Hawthorn, VIC 3122, Australia \and
     Observatories of the Carnegie Institution of Washington,
     Pasadena, CA 91101, USA \and
     The Niels Bohr Institute, Juliane Maries Vej 30, 2100 Copenhagen, Denmark\\
   }

   \offprints{Jesus Varela,\\ \email{jesus.varela.lopez@gmail.com}}
   \date{\today}


  \abstract
  {This is the second  paper of a  series devoted to the {\textit{WIde
        Field Nearby Galaxy-cluster Survey}} (WINGS).  WINGS is a long
    term project which is gathering wide-field, multi-band imaging and
    spectroscopy  of  galaxies  in a   complete  sample  of   77 X-ray
    selected, nearby   clusters ($0.04<z<0.07$)  located  far from the
    galactic plane ($\left|b\right|\geq20^o$).  The  main goal of this
    project is to establish a local reference for evolutionary studies
    of galaxies and galaxy clusters.}
  {This paper presents the optical (B,V) photometric catalogs of the
    WINGS sample and describes the procedures followed to construct
    them.  We have paid special care to correctly treat the large
    extended galaxies (which includes the brightest cluster galaxies)
    and the reduction of the influence of the bright halos of very
    bright stars.}
  {We have constructed photometric catalogs based on wide-field images
    in B and  V bands using \sext.  Photometry  has been performed on
    images  in which large  galaxies  and halos  of bright stars  were
    removed    after modeling   them    with   elliptical   isophotes. 
    }
    {We publish deep optical photometric catalogs (90\%
        complete at $V\sim21.7$, which translates to $\sim M^*_V+6$ at
        mean redshift), giving positions, geometrical parameters, and
      several total and aperture magnitudes for all the objects
      detected.  For each field we have produced three catalogs
      containing stars, galaxies and objects of ``unknown''
      classification ($\sim 16\%$).  From simulations we found that
      the uncertainty of our photometry is quite dependent of the
      light profile of the objects with stars having the most robust
      photometry and de Vaucouleurs profiles showing higher
      uncertainties and also an additional bias of $\sim -0.2^m$.
    
      The star/galaxy classification of the bright objects ($V<20$)
      was checked visually making negligible the fraction of
      misclassified objects.  For fainter objects, we found that
      simulations do not provide reliable estimates of the possible
      misclassification and therefore we have compared our data with
      that from deep counts of galaxies and star counts from models of
      our Galaxy.  Both sets turned out to be consistent with our data
      within $\sim 5\%$ (in the ratio galaxies/total) up to $V\sim
      24$.
       
      Finally, we remark that the application of our special procedure
      to remove large halos improves the photometry of the large
      galaxies in our sample with respect to the use of blind
      automatic procedures and increases ($\sim$16\%) the detection
      rate of objects projected onto them.} {}

   \keywords{Galaxies : Clusters : General - Catalogs}
               \titlerunning{WINGS II: Deep Optical Catalogs}
   \maketitle
%

\section{Introduction}

%
%

Clusters of galaxies are privileged systems to study, among others,
two basic problems. On one hand, they (or a substantial part of them
at least) are the largest gravitationally bound structures and as
such, physical entities whose properties must be explained by
cosmological theories. On the other hand, clusters are made of
hundreds or even thousands of galaxies in high density regions whose
properties can be studied and straightforwardly compared and their
evolution analyzed.

Clusters of galaxies were first detected as marked overdensities in
the projected number of galaxies.  Even if today it has been
recognized that galaxies represent only a small fraction of the total
mass of the clusters, they still are a fundamental tool to study
cluster properties since they are usually much more easily detected
and measured than the X-ray emitting intracluster gas or the more
evasive non-baryonic component.

Clusters of galaxies also have been widely used to study the evolution
of galaxies in dense environments.  Some of the first clear evidence of
evolution of  the galaxies came from  the study of  the populations of
galaxies in clusters made by \citet{Butcher1978a}. They found that the
fraction of blue galaxies was higher in clusters at $z\gtrsim0.4$ than
in nearby clusters, which they interpreted as the result of the aging of
spiral    galaxies after  losing  their   gas  supply and,  therefore,
diminishing their star formation rates.  In parallel,
\citet{Dressler1980a} showed   that the central,   denser  parts of 
clusters  are mainly populated  by  early  type galaxies.  Since then,
several works have discussed  the  idea that morphological content  in
clusters of galaxies changes along the Hubble time
\citep{Dressler1997,    Fasano2000,     vanDokkum2000,      Lubin2002,
Postman2005}.

Two are the main reasons that make clusters useful to study the
evolution of galaxies. First, galaxies in clusters can be considered
as being at the same distance since usually this distance is much
greater than the linear dimensions of the clusters. Therefore, knowing
the redshift of a small subsample of the cluster is enough to
(statistically) know the distance (once background correction it is
allowed for) of hundreds or even thousands of galaxies. Of course, this
highly increases the statistics when analyzing the properties of the
galaxies. Second, clusters of galaxies can be detected up to high
redshifts using different techniques that are sensitive to local
enhancements of the galaxy density or through the X ray emission from
the intracluster medium.

Using the first all sky survey~\citep[The National Geographic
Society-Palomar Observatory Sky Survey,][]{Abell1959} the first
systematic catalogs of clusters of galaxies were
constructed~\citep{Abell1957,Zwicky1963}. However, it was only in the
1970s when clusters of galaxies started to be used as laboratories to
study the properties of the galaxies they contain and how their
evolution is affected by the environment ~\citep{Gunn1972,Oemler1974}.
From this point on, the interest in clusters of galaxies at
always higher redshift continuously grew and nowadays, specially
thanks to the Hubble Space Telescope, clusters at $z\gtrsim1$ have
been studied in detail~\citep[e.g.  RDCS J1252 2927 at $z=1.235$;
][]{Postman2005} and using different techniques \textit{protoclusters}
have been detected at $z>2$ \citep[e.g.][]{Steidel2000,Kurt2004}.

To produce a correct interpretation of the observations at different
redshifts a good knowledge of the cosmic variance of the properties of
clusters and galaxies at each redshift is needed, to ensure that the
changes observed with distance are statistically significant.
Paradoxically, as we discussed in \citet[][hereafter Paper
I]{Fasano2006}, there is a relative lack of knowledge of the
  properties of the clusters of galaxies in the local Universe at
$z\sim0$ and the comparison of higher redshift clusters is always done
with single clusters such as Virgo, Coma or Fornax. Indeed, for more
than 20 years the most complete study of galaxies in nearby clusters
of galaxies has been that by \citet{Dressler1980a}.  Dressler's work
was based on data extracted from photographic plates.  Several
programs have been set up to continue that work with modern tools and
techniques. Each of them addresses a particular aspect or problem and,
therefore, we still lack a general local reference for evolutionary
studies.  Thus, the ESO Nearby Abell Cluster
Survey~\citep[][ENACS]{Katgert1996,Biviano1997} includes spectroscopy
of galaxies in a number of clusters but the imaging is not as deep as
that of Dressler's work.  The Las Campanas/ATT Rich Cluster
Survey~\citep{Ohely1998,Pimbblet2001} is deeper but includes only 20
clusters.  In recent years, large sky surveys as the
2dF~\citep{DePropris2002} and, particularly, the
SDSS~\citep{Goto2002,Bahcall2003,Miller2005} have been sources for the
compilation and analysis of large samples of clusters of galaxies.
However, while the 2dF is only spectroscopic the SDSS, which covers a
very large area, is not deep enough to study the faintest part of the
luminosity function.  More recently, the NOAO Fundamental Plane Survey
~\citep{Smith2004} started to fill this lack of data, however its main
goal is not the study of the evolution of galaxies and clusters but
the large scale velocity fields using the Fundamental
Plane.\footnote{A more detailed comparison from the spectroscopic
  point of view will be made in \citet{Cava2008}.}

In this context, the \textit{WIde-field Nearby Galaxy-clusters Survey}
\citepalias{Fasano2006}  has   been presented with   the   specific goal of
sampling the properties of  clusters and galaxies  in clusters  in the
local  Universe. This  means to  establish  both the  average and  the
variance of  the properties   of clusters  and  of the   galaxies they
contain.

%
%

Briefly, WINGS is a long term multiwavelength project based on deep
optical (B,V) wide field images ($\sim35$'$\times 35$') of 77 fields
centered on nearby clusters of galaxies selected from three X-ray flux
limited samples compiled from ROSAT All-Sky Survey data (BCS,
\citealt{Ebeling1996}; eBCS, \citealt{Ebeling1998}; XBACs,
\citealt{Ebeling2000}).  The selected redshift range
(\mbox{$0.04<z<0.07$}) was set to balance a wide linear field of view
($\sim 1.4\,$Mpc$\,\times\,1.4\,$Mpc at $z\sim0.04$) and a high
spatial resolution (1.34\,kpc/'' at $z\sim0.07$).  To reduce the
effects of galactic extinction in our analysis, only clusters located
far from the galactic equator \mbox{($\left| b\right| \ge 20^o$)} were
kept.  The observations were carried out using the wide field cameras
of two telescopes: the 2.5m Isaac Newton Telescope (WFC@INT) and the
MPG/ESO-2.2m telescope (WFI@ESO).  The optical data were complemented
with spectroscopic follow up of a subsample of 48
clusters~\citep{Cava2008} using WYFFOS@WHT (\mbox{$\lambda\ 
  $range=3800-7000 \AA}, \mbox{$\lambda\ $resolution=3\AA}) and
2dF@AAT (\mbox{$\lambda\ $range=3600-8000 \AA}, \mbox{$\lambda\ 
  $resolution=6\AA}). Also, we obtained J and K imaging of 32 WINGS
clusters with WFCAM@UKIRT~\citep{Valentinuzzi2008} and we are
collecting U/H$_\alpha$ band imaging of WINGS clusters with wide-field
cameras at different telescopes (INT, LBT, Bok), that will be used to
analyse  the stellar masses and of star formation.

\citetalias{Fasano2006} started the series devoted to the analysis of
the optical data within the so called WINGS-OPT subproject.  It
presented the criteria followed to construct the initial sample of
clusters and described the optical observations (B and V bands) and
their photometric and astrometric quality.

\begin{figure*}
\def\scf{0.9}
  \centering
  \includegraphics[scale=\scf]{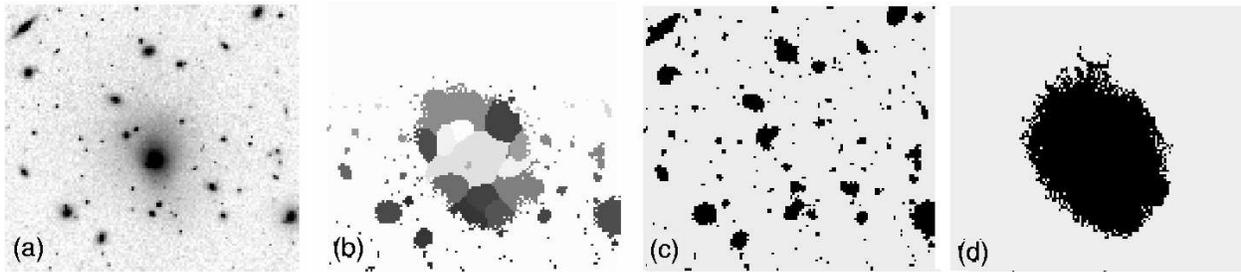}
  \caption{Example of the problems  of \sext\ to
    produce the correct segmentation of the extended galaxies. From
    left to right: (a) BCG of cluster A193 in the V band image; (b)
    segmentation performed by \sext\ without subtracting the galaxy;
    each color represents those pixels assigned by \sext\ to different
    objects ; (c) segmentation of the remaining objects after
    subtraction of the BCG; (d) final segmentation of the BCG.}
  \label{fig:A193bcg}
\end{figure*}

%
%

This is the second paper in the series regarding the optical
photometry of the WINGS project and it is devoted to the release of
the catalogs containing the basic photometric parameters of all the
objects found in the fields of 77 nearby clusters of galaxies. Also,
it describes the procedures that were followed to construct the
catalogs (Section ~\ref{sec:Preprocess}).  The process includes a
preliminary treatment of the images
({$\S$}~\ref{sec:PreprocessImages}) that has allowed us to improve the
photometry of many galaxies, especially the largest ones.  In
Section~\ref{sec:Catalogs}, catalogs are presented and their overall
quality is checked including their photometric quality
({$\S$}~\ref{sec:PhotErrors}), completeness and star/galaxy separation
({$\S$}~\ref{sec:Completeness}).  A final summary will be found in
Section~\ref{sec:Summary}.

%
%

Throughout  this paper   we  will   use   a  cosmological model   with
parameters:       \mbox{$H_0=75$         km$\,$s$^{-1}\,$Mpc$^{-1}$},\
$\Omega_M=0.3$ and $ \Omega_\Lambda=0.7$ \footnote{Be aware that in
  \citetalias{Fasano2006} we used instead $H_0=70$         km$\,$s$^{-1}\,$Mpc$^{-1}$,
however this change doesn't modify the results showed there}.


\section{Detection, basic photometry and star-galaxy classification\label{sec:Preprocess}}

The whole process of source detection, computation of their basic
photometric parameters and star-galaxy classification was performed
using \sext~\citep{Bertin1996}.  \sext\ is a very efficient tool to
find sources in large format images.  Among its characteristics the
most interesting are its ability to separate blended sources and to
estimate the probability of an object of being a star or an extended
source.  However, both processes need to be fine-tuned by input
parameters that are fixed for each single image. To improve the final
outcome of \sext\ a preliminary treatment of the images was done
before running the source detection program as well as a final
interactive checking of the star-galaxy classification, as described
in the following.

\subsection{Preliminary image treatment\label{sec:PreprocessImages}}

In images like ours in which the ranges of size and brightness are
quite large, it is impossible in practice to find a single set of
values of \sext's input parameters that could work adequately for all
the objects.  The extreme situation is when a large galaxy is
contaminated by a number of small projected sources as  is the case
of the brightest central galaxies (BCGs) of our sample of clusters.
This issue is illustrated in Fig.\ref{fig:A193bcg} with the BCG of the
cluster A193 (panel (a)).  The segmentation\footnote{The segmentation
  is the way in which a program of source detection assigns pixels to
  each object.}  of the central galaxy and the objects projected onto
it is shown in panel (b).  Clearly, the program has erroneously
assigned pixels of the large galaxy to the small projected objects
producing poor photometry of the large galaxy as well as of the
objects projected onto it. Something similar happens with very bright
stars whose extended halos affect the photometry of the nearby
objects.

To minimize the effect of such large halos in the photometry of close
objects as well as to improve the photometry of the extended galaxies
themselves we have developed a custom-made procedure.  It consists of
first the modeling and removal of these halos before running \sext\ on
the image. This improves the photometry of the small projected
objects.  In a second step, an image is constructed containing only
the extended galaxies and the photometry is performed on it with
\sext. In this last image the pixels of the projected small objects
are replaced by the values of the models and the rest of the pixels
are left unchanged. This replacement reduces the contamination of the
projected objects in the photometry of the large galaxies. An
additional advantage of the procedure is that the removal of the
extended halos also improves the determination of the global
background map.


%
%

In the following we  give a brief description  of the process that  is
explained in more detail in Appendix~\ref{sect:ap1}.

The process starts with the computation of a first background map
which is subtracted from the image.  Then, the problematic objects
(galaxies and stars with extended halos) are located.  For each of
these objects, a mask of the projected objects is constructed and
elliptical isophotes are fit\footnote{To achieve a successful
    fit a preliminary 'masking' procedure is needed of the objects
    that appear projected onto the halos. See Appendix~\ref{sect:ap1}
    for more details.}. The resulting fit is used to construct a
model of the halo that is afterwards subtracted from the original
image (i.e.  the image before the background subtraction).  After
doing that with all the selected objects, the resulting image without
the large halos is used to refine the background map as well as the
masks of the smaller objects.  With the new background map and masks
the process is repeated to improve the final results.  It is found that
one iteration is enough to achieve sufficiently accurate photometry.

The last step is the construction of a complementary image containing
only the previously removed galaxies (of course, the bright stars are
also avoided in this image).  It is important to note that the
photometry of the large galaxies is not performed on the models but
rather using the original pixels.  The models are used only those
regions occupied by projected galaxies or by interchip regions.

At the end of the whole process, which is run in V band as well as in
B band, the output is two background subtracted images in each
band. One of the images contains all the objects except the largest
ones and the other one only the extended galaxies that were removed
from the first image.

The improvement of the procedure is illustrated in the last two panels
of Fig.\ref{fig:A193bcg} in   which we show  the  segmentations of the
projected objects (panel    (c)) and of   the  BCG (panel  (d))  after
applying our procedure.

%

\begin{figure*}
  \def\scf{0.2}
  \centering
  \includegraphics[scale=\scf]{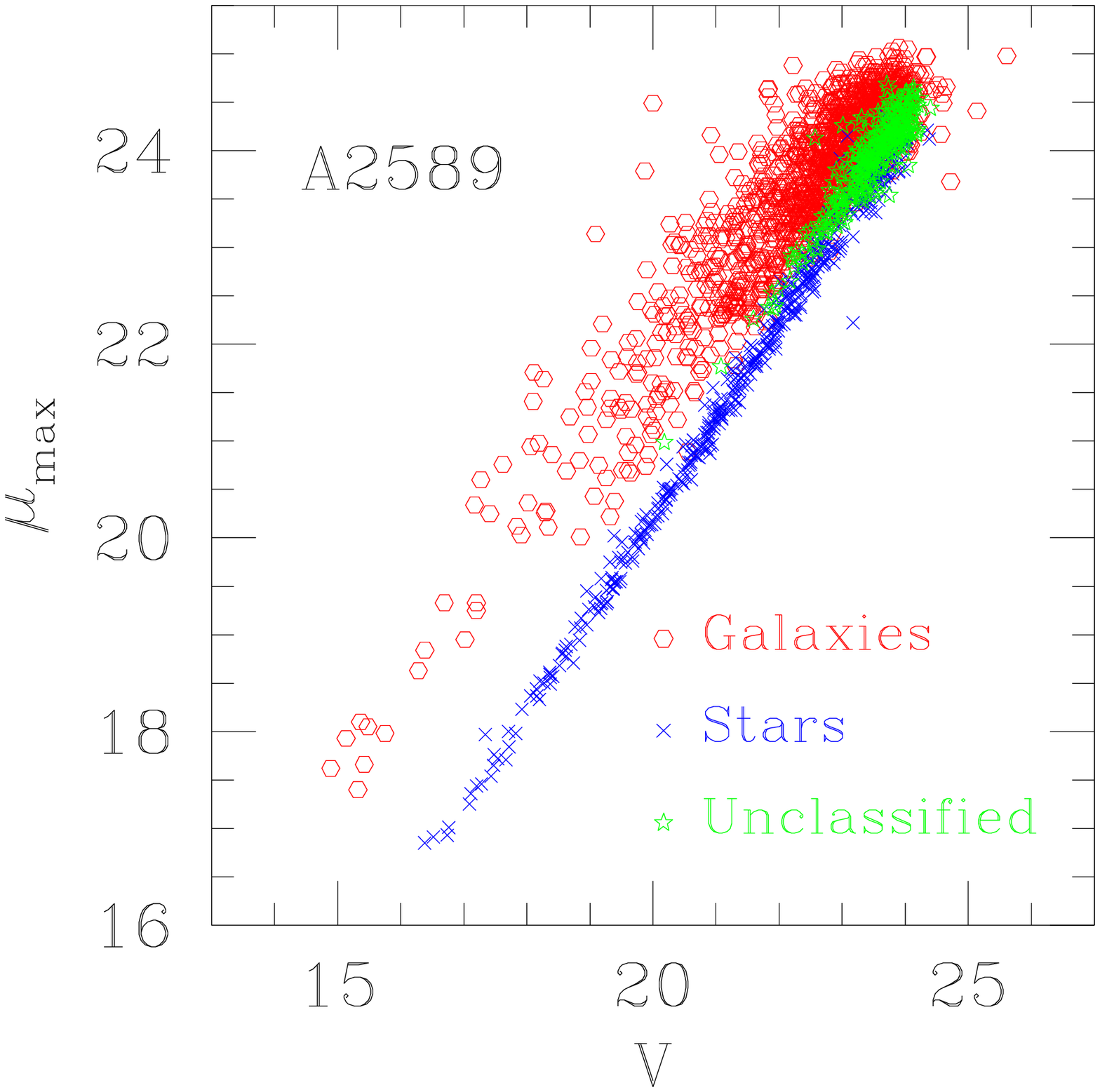}
  \includegraphics[scale=\scf]{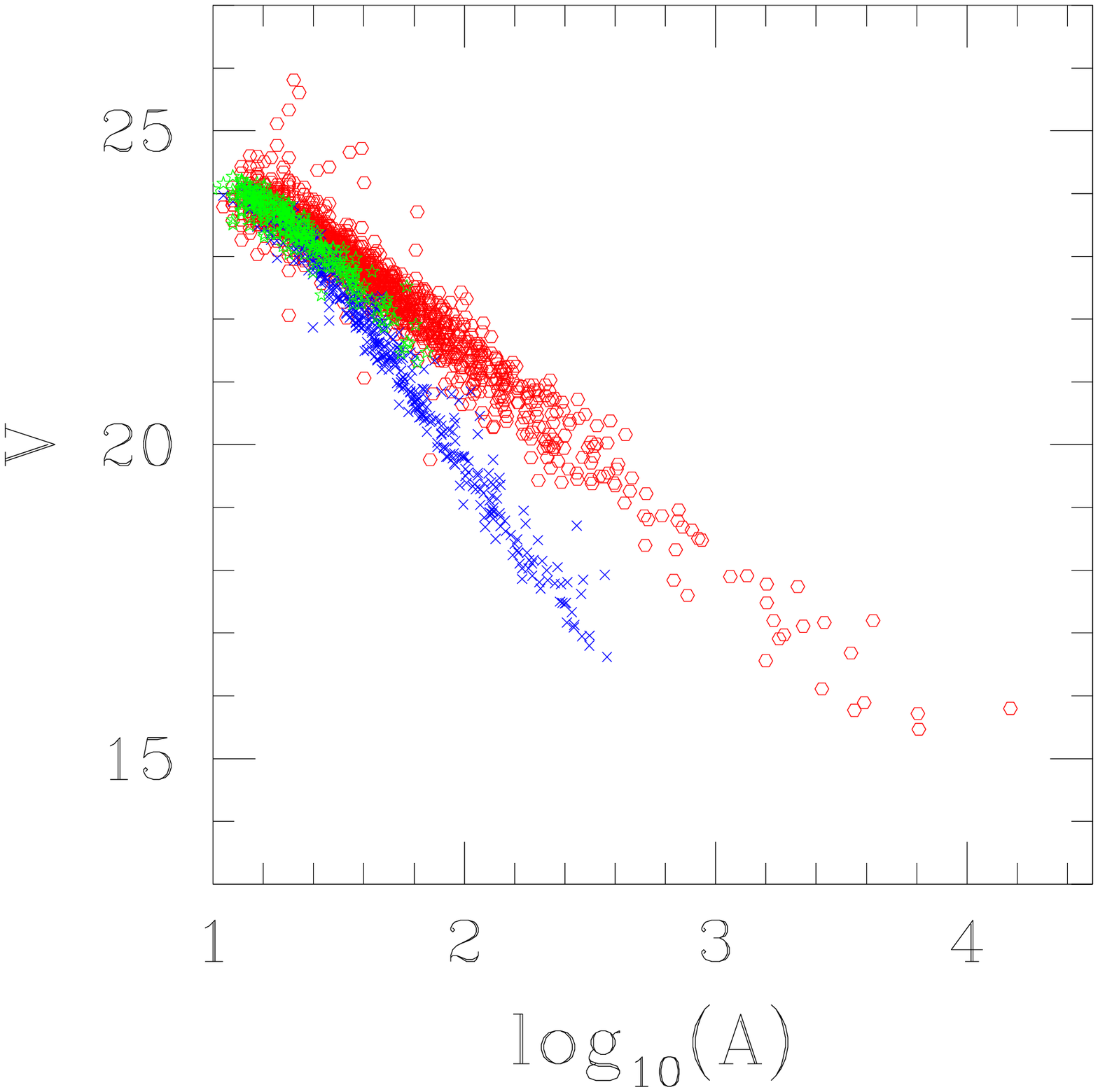}
  \includegraphics[scale=\scf]{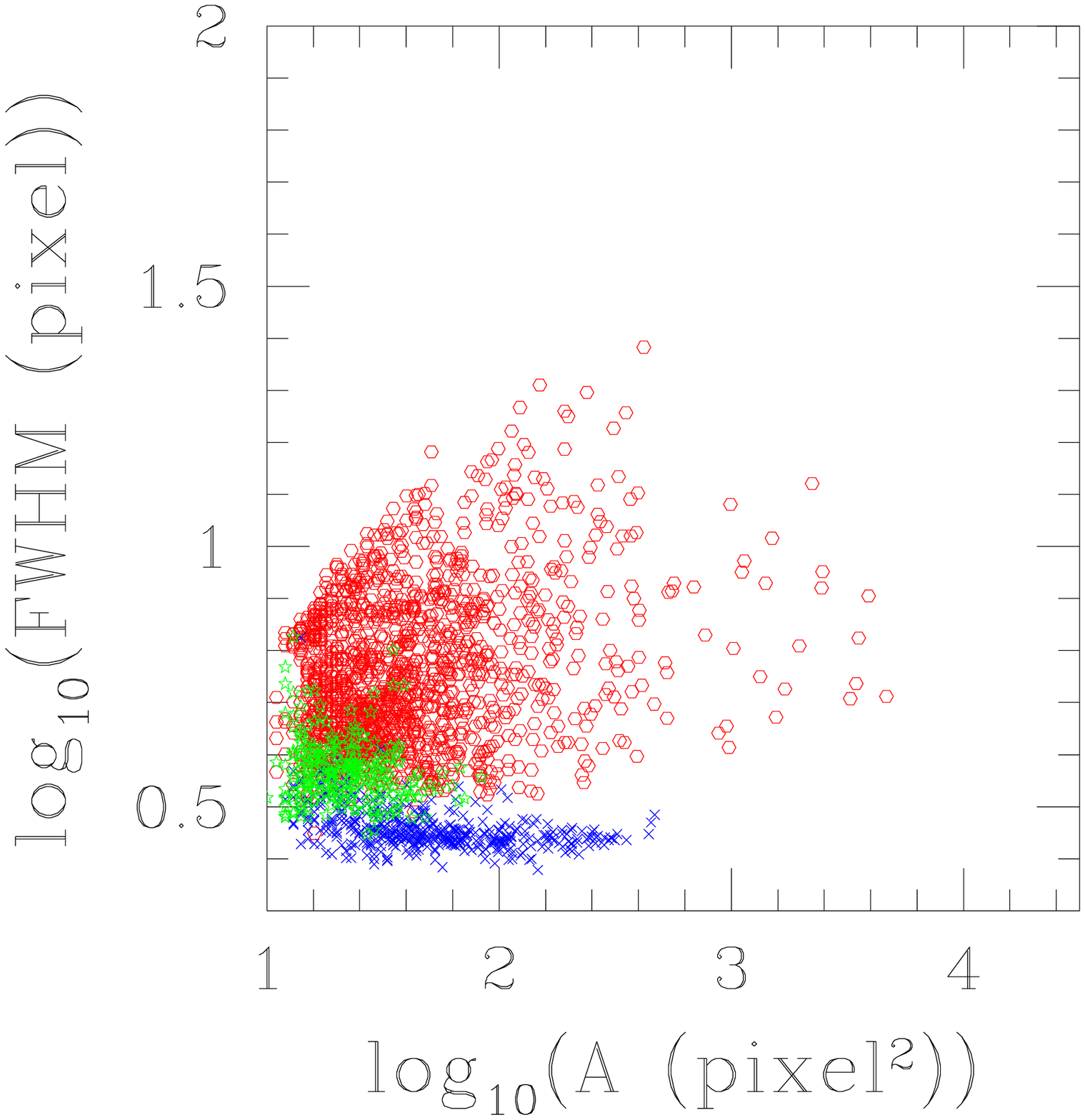}
  \includegraphics[scale=\scf]{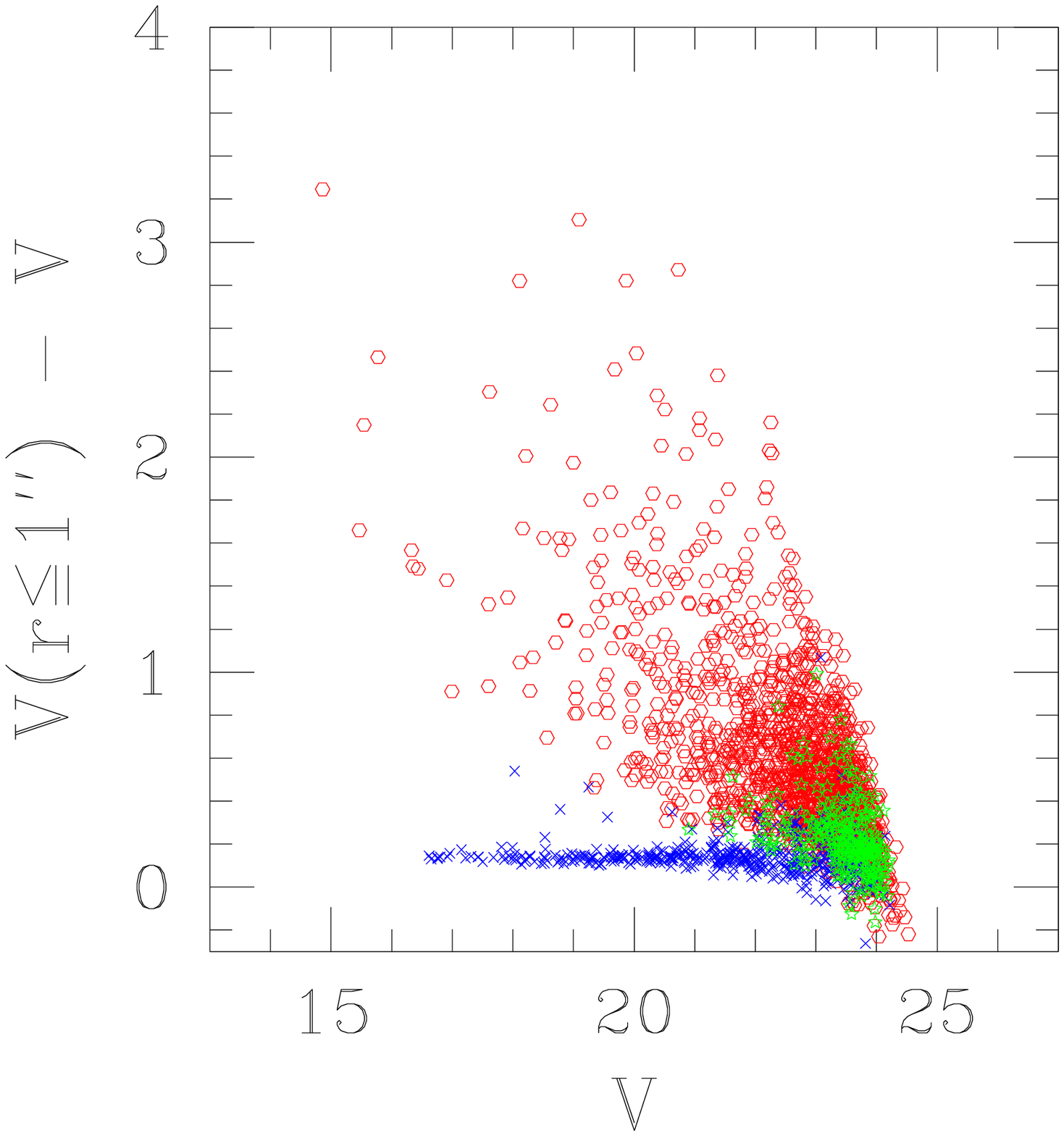}

  \includegraphics[scale=\scf]{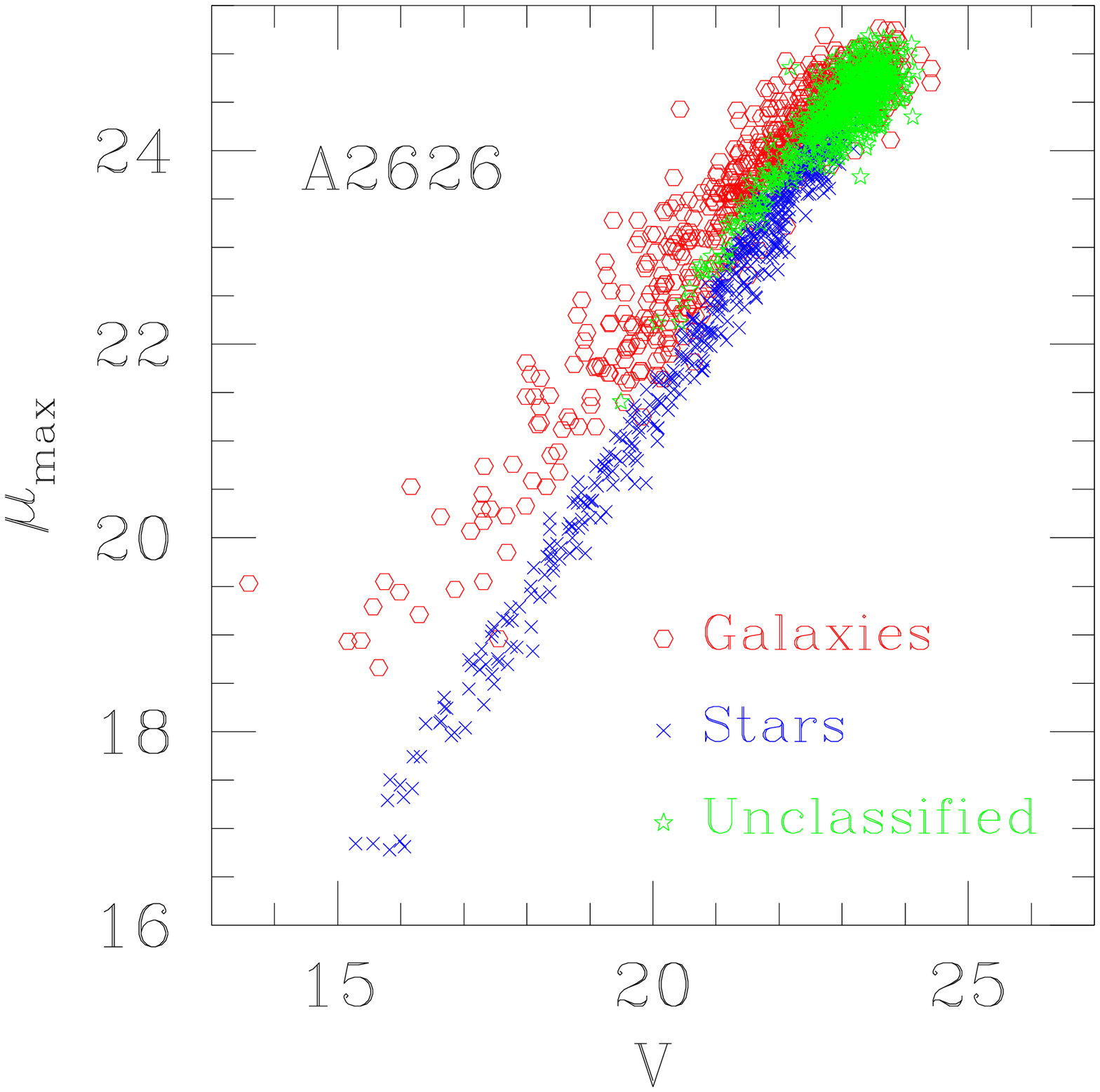}
  \includegraphics[scale=\scf]{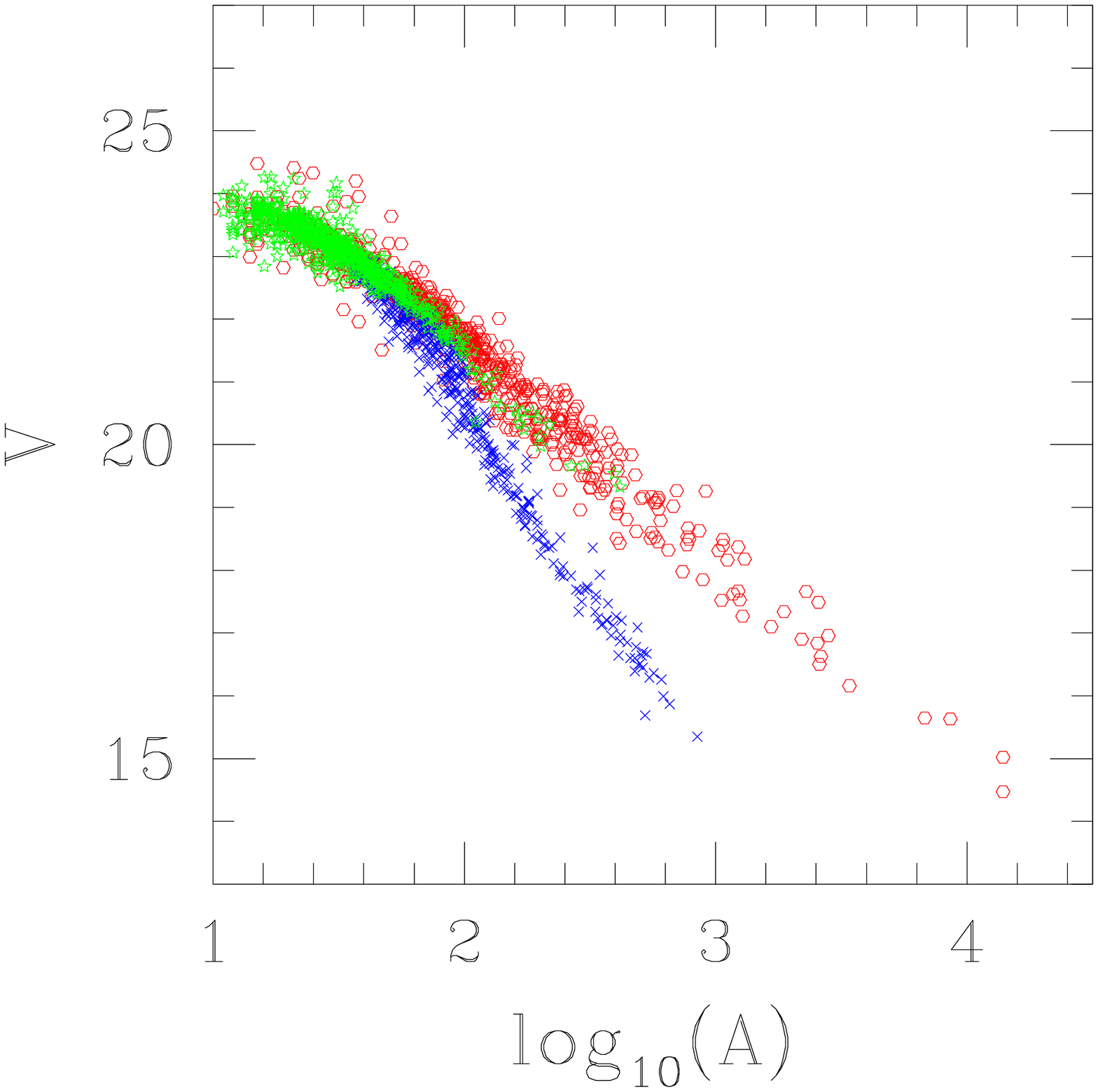}
  \includegraphics[scale=\scf]{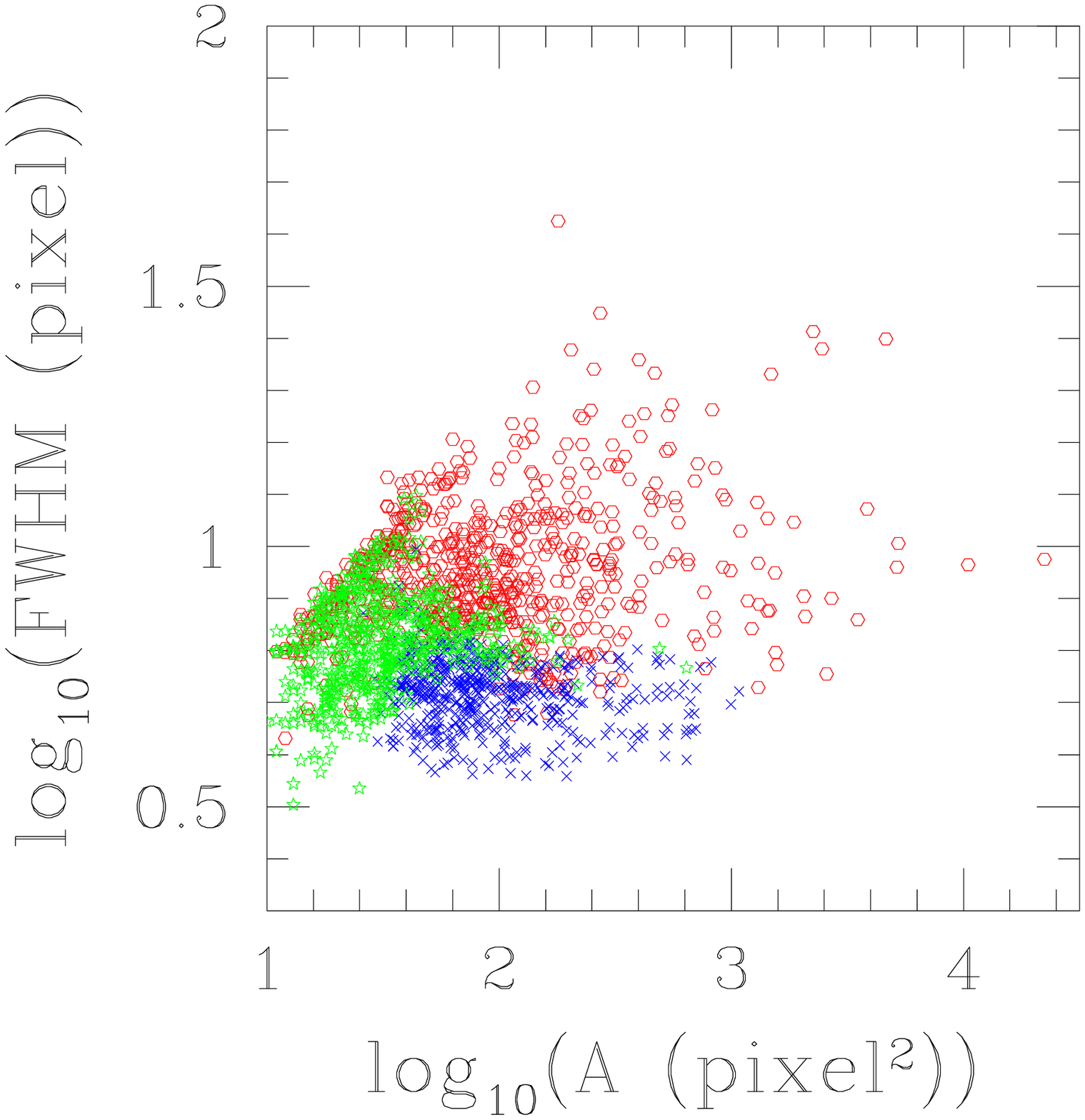}
  \includegraphics[scale=\scf]{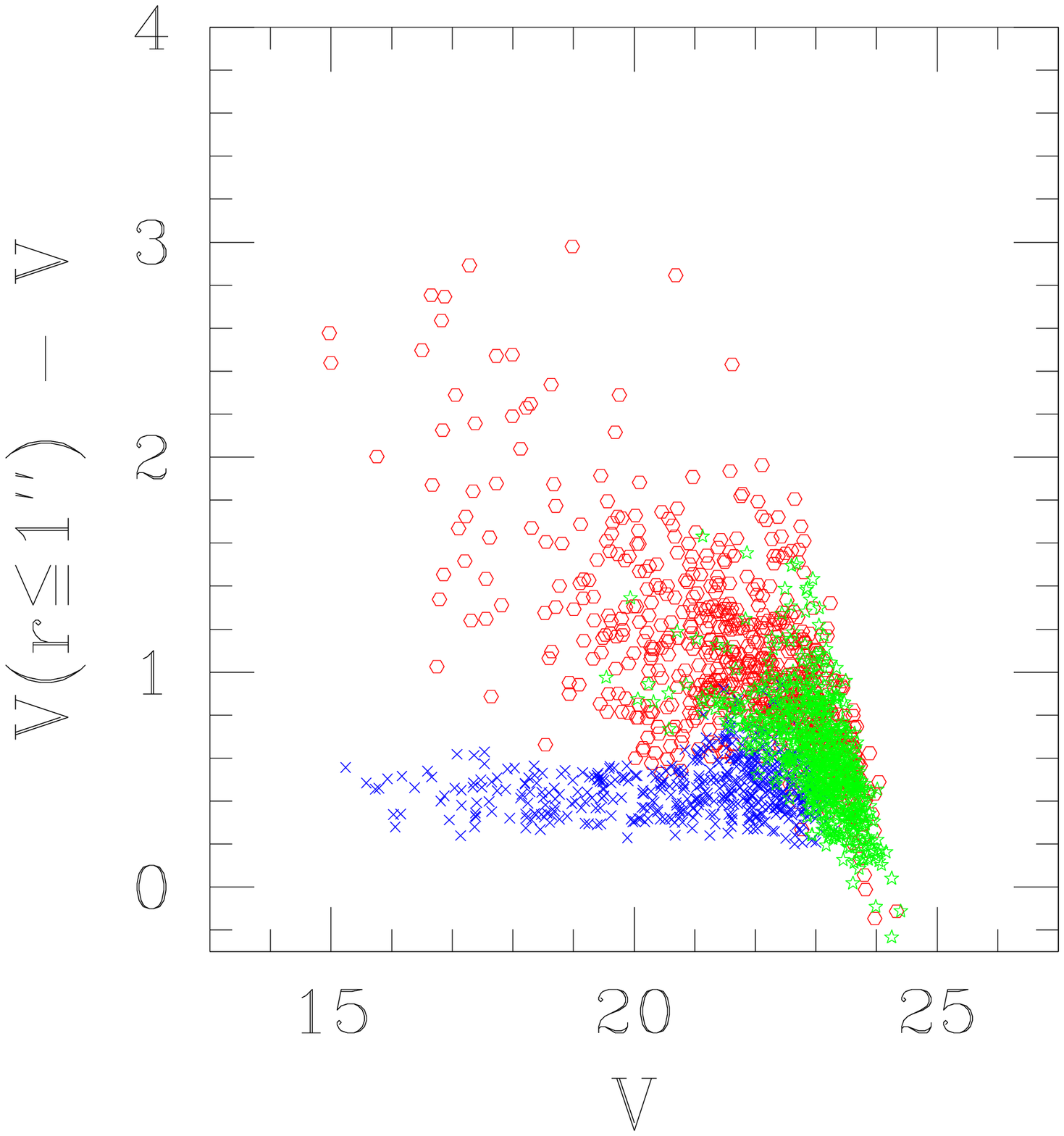}
  \caption{Four different kinds of diagrams which can be used to
    perform a star/galaxy separation.   From left to right: \mumax-V ;
    V-$\log(Area)$  ;       $\log(\fwhm)-\log(Area)$;     $V-
    (V(r\le 1")-V)$.     The data come  from  two  fields with quite
    different  seeing   conditions: A2589 ($\fwhms=0.\arcsec85$; upper
    panels) and    A2626 ($\fwhms=1.\arcsec64$;  lower   panels).  Red
    squares    are  galaxies, blue     crosses stars  and   green dots
    unclassified objects.  For  better  visualization, only $20\%$  of
    the objects have been plotted.}
  \label{fig:Vmumax}
\end{figure*}

\subsection{Detection and basic photometry}

To construct the  photometric catalogs, the  detection of sources  and
the basic  photometry relies  on  \sext\ and more  specifically on the
modified              version            \sext\     2.3.2           by
G. Morrison\footnote{http://www.cfht.hawaii.edu/\~{}morrison/home/SExtractor.html}
following  the changes   initially  made  by  B.  Holwerda for  \sext\
v.2.2.2.
 These versions  have the advantage  of computing additional parameters
related to  the  light  distribution in  the  objects such   as  the
concentration,      the      contrast            and               the
asymmetry~\citep{Abraham1994,Abraham1996}.

The detection of sources, the determination of their positions and
geometrical parameters as well as the star/galaxy classification was
done on the V band image.  Since we had images from two different
cameras with different pixel scales, instead of setting a single
detection threshold per pixel for both cameras we opted for setting
the same detection threshold per square arcsec. The advantage of this
approach is that with similar observational conditions the limiting
surface brightness is independent of the pixel size.  The disadvantage
is that in the images with lower detection threshold per pixel (in our
case the WFI@ESO) the rate of spurious detections at low brightness is
higher.  However, since we were interested in the bright part of the
luminosity function, where the galaxies of the cluster outnumber those
of the background, it was desirable to perform the photometry of the
galaxies within similar surface brightness limits.  In order to
balance photometric depth (i.e.  detection of low surface brightness
galaxies) with a low rate of spurious detections, the detection
threshold was set to 4.5\,\sbg/arcsec$^2$ (\sbg\ is the standard
deviation of the background signal) which corresponds to
1.5\,\sbg/pixel for the WFC@INT and 1.07\,\sbg/pixel for the WFI@ESO.
Given the typical values of \sbg\ found in our images these limits
translate in a detection limit of $\mu_{Threshold}(V)\sim 25.7\,
$mag/arcsec$^2$.  As a comparison, at the same signal to noise level
the images from the SDSS\footnote{The following values have been
  calculated from the image identified by
  (run,rerun,camcol,field)=(1889,40,3,105) which is the one taken in
  most favorable conditions} in g band reach $\mu_{Threshold}(g)<
25.2\, $mag/arcsec$^2$ while those in r band reach
$\mu_{Threshold}(r)< 24.7\, $mag/arcsec$^2$.

\sext\ was run twice with the B band images using both the
single-image and the dual-image modes\footnote{In single-image mode
  the detection and the photometry are done using the same image.  In
  dual-image mode the detection is done in one reference image and the
  photometry of the detected objects is done in a second image.}.  The
catalog resulting from running \sext\ in single-image mode allowed us
to reduce the number of spurious detection because only objects with
detections in both bands (separated by less than
1\farcs67\,=\,5\,pixel\,(WFC@INT)\,=\,7\,pixel\,(WFI@ESO)) were kept.
However, to obtain correct color indexes of the galaxies we performed
a second run of \sext\ on the B band image using the dual-image mode
in which the V band image was set as the detection image.  This procedure
ensured that the measurements of each galaxy in both filters were done
in the same regions.

\sext\ was run without using weighted images because preliminary tests
showed that the photometry was more robust when not using weighting
images\footnote{Nowadays, the use of weighted images is widely
    used and a safe procedure, however, at the moment of constructing
    the catalogs we saw that the use of the weighted images increased
    the uncertainties in the magnitudes so we opted for not using
    them.}. As a consequence, photometric errors computed by \sext\ 
are unrealistically small. The estimation of these errors is explained
in Section~\ref{sec:PhotErrors}.

To avoid the influence of the interchip regions, the pixels in these
regions were set to zero and the objects falling within a band of
several pixels to one of the edges of the chips\footnote{The bands
  were constructed interactively to ensure that the problematic
  regions were avoided.}  were excluded from our catalogs.

The saturated stars\footnote{The exposure time of the single exposures
  were set to avoid the saturation level for galaxies so no galaxy was
  found to reach such a level.} were not included in the catalogs
because of the uncertainty in their photometry.

Once having obtained the photometry in both bands, we applied the
color term correction of the flux calibration.  This was done using
the color index (B-V) measured in apertures of R=5\,kpc, except for
objects with smaller linear radius for which smaller apertures
(R=2\,kpc) were used.

For each field, \sext\ was run on the images obtained for both the
bright and faint galaxies (see Sect~\ref{sec:PreprocessImages}) and
the final catalog was obtained by merging the two catalogs.

\subsection{Star-galaxy classification\label{sec:stargal}}

The  star-galaxy classification was made  relying  upon the stellarity
index  (\texttt{CLASS\_STAR}) computed   by \sext\ which   ranges from
0~(galaxies)   to  1~(stars). We decided   to   make  a rather  robust
selection in two groups,``stars''  and ``galaxies'', with  the highest
probability of being correct. Objects  not fitting those criteria were
left in a  third  group of ``unknown'' classification.  Practically we
imposed the following criteria:

\begin{tabular}{ll}
\textbf{Stars}    & $\texttt{CLASS\_STAR} \ge 0.8$  \\
\textbf{Galaxies} & $\texttt{CLASS\_STAR} \le 0.2$  \\
\textbf{Unknown}  & $0.2<\texttt{CLASS\_STAR} <0.8$ \\
\end{tabular}

\begin{table*}[t]
  \caption{Example of entries in one of the photometric catalogs.
    \label{tab:Catalogs}
    }
    \tiny
    \begin{tabular}{|*{14}{c}}
      \hline
      \multicolumn{1}{|c}{ID} & $\alpha_{Bary}$ & $\delta_{Bary}$ & 
      $\alpha_{Peak}$ & $\delta_{Peak}$ & 
      Area &  $r_{Kron}$ & FWHM & b/a & PA & \texttt{FLAG} &
      \texttt{S.I.} & &\multicolumn{1}{c|}{}\\
      \multicolumn{1}{|c}{} 
       & (deg) & (deg) & (deg) & (deg) & (arcsec$^2$) & (\arcsec) 
       & (\arcsec) & & (deg) & \multicolumn{4}{c|}{} \\
      \hline\hline
      WINGS J004135.4-090100.5 & 10.39751 & -9.01681 & 10.39751 & -9.01682 & 17.39 & 1.49 &
      6.26 & 0.73 & 63 & 19 & 0.02  
      & \ldots \\
      WINGS J004106.0-090104.2 & 10.27495 & -9.01783 & 10.27500 & -9.01781 & 41.86 & 0.83 &
      2.03 & 0.75 & 31 & 0 & 0.03  
      & \ldots \\
      \hline\hline
    \end{tabular}
\newline
  \scriptsize
  \begin{tabular}{*{14}{c}}
    \multicolumn{14}{c}{} \\
    \multicolumn{14}{c}{} \\
    \hline
    \multicolumn{1}{|c}{}
    & \multicolumn{12}{c}{V band} &     \multicolumn{1}{c|}{}\\
    \cline{2-13}
    \multicolumn{1}{|c}{}
     & Conc. & Contr. & $\mu_{max}$ & \texttt{MAG\_ISO} &\texttt{MAG\_ISOCOR} & \texttt{MAG\_AUTO} &
    \texttt{MAG(2Kpc)} & \texttt{MAG(5Kpc)} & \texttt{MAG(10Kpc)} & 
    \texttt{MAG(1\farcs6)} & \texttt{MAG(2\farcs0)} &
    \texttt{MAG(2\farcs16)} & \multicolumn{1}{c|}{}\\ 
    \hline\hline
    \ldots &
   0.384 & 0.543 &
    22.97 & 20.71 & 20.44 & 19.86 & 
    21.43 & 20.63 & 20.10 & 
    22.53 & 22.19 & 22.09 
    & \ldots \\
    \ldots &
    0.460 & 0.777 &
    21.03 & 18.58 & 18.55 & 18.57 &
    19.11 & 18.60 & 18.51 & 
    20.48 & 20.09 & 19.97 
    & \ldots \\
    \hline\hline
  \end{tabular}
\newline
  \scriptsize
  \begin{tabular}{*{11}{c}|}
    \multicolumn{10}{c}{} \\
    \multicolumn{10}{c}{} \\
    \hline
    \multicolumn{1}{|c}{}
    & \multicolumn{9}{c}{B band}&  \multicolumn{1}{c|}{} \\
    \cline{2-10}
    \multicolumn{1}{|c}{}
    & \texttt{MAG\_ISO} &\texttt{MAG\_ISOCOR} & \texttt{MAG\_AUTO} &
    \texttt{MAG(2Kpc)} & \texttt{MAG(5Kpc)} & \texttt{MAG(10Kpc)} & 
    \texttt{MAG(1\farcs6)} & \texttt{MAG(2\farcs0)} & 
    \multicolumn{1}{c}{\texttt{MAG(2\farcs16)}}  &  \multicolumn{1}{c|}{} \\ 
    \hline\hline
    \ldots &
    21.41 & 20.46 & 20.00 & 
    22.18 & 21.09 & 20.21 & 
    23.37 & 23.01 & 22.90 
    & \\
    \ldots &
    19.60 & 19.48 & 19.58 &
    20.16 & 19.61 & 19.47 & 
    21.58 & 21.18 & 21.05 
    & \\
    \hline\hline
  \end{tabular}
\begin{flushleft}
  ID: Object internal identification.\\
  ($\alpha_{Bary}$,$\delta_{Bary}$) : Equatorial coordinates (J2000.0) of the barycenter.\\
  ($\alpha_{Bary}$,$\delta_{Bary}$) : Equatorial coordinates (J2000.0) of the brightest pixel.\\
  Area: Area above the detection threshold.\\
  $r_{Kron}$ : Kron radius used to computed the \texttt{MAG\_AUTO} magnitude.\\
  FWHM: Full width at half maximum assuming a Gaussian core.\\
  b/a : Axis ratio.\\
  PA : Position angle of the major axis (North=0\degr, counter-clockwise).\\
  \texttt{FLAG} : \sext's \texttt{FLAG}\\
  S.I. : \sext's stellarity index \texttt{CLASS\_STAR}\\
  $\mu_{max}$ : Surface brightness of the brightest pixel.\\
  Conc. : Abraham's concentration index, measured as  the  ratio between  the  flux in the
  central 9\% of the pixels and the total flux.\\
  Contr. : Abraham's contrast index, measured  as the  ratio  between the flux in the
  brightest 30\% of the pixels and the total flux.\\
  \texttt{MAG\_ISO} : \sext's isophotal magnitude.\\
  \texttt{MAG\_ISOCOR} : \sext's isophotal corrected magnitude.\\
  \texttt{MAG\_AUTO} : \sext's Kron aperture magnitude.\\
  \texttt{MAG(2Kpc),   MAG(5Kpc),  MAG(10Kpc)}   :  Magnitudes  within
  apertures of radius 2Kpc, 5Kpc and 10Kpc, respectively, measured at
  the clusters' redshift.\\
  \texttt{MAG(1\farcs6),  MAG(2\farcs0),  MAG(2\farcs16)} : Magnitudes
  measured within  apertures   of diameter equal  to  1\farcs6, 2\farcs0,
  2\farcs16, respectively.
\end{flushleft}
\end{table*}
\normalsize



In spite  of the robustness  of the criteria  that we used,  the final
catalogs were   checked  for possible  star/galaxy misclassifications.
Since the number of objects was too large for any individual analysis,
we opted for using    several   plots of different  combinations    of
parameters  to    detect  these  possible   misclassifications.     In
Fig.~\ref{fig:Vmumax} we show  some of the  combinations of parameters
that separate stars from galaxies:
\begin{itemize}
\item $V-\mumax$,   where \mumax\  is  the  surface brightness   of the
  brightest pixel in  an object and $V$  is the total V  band apparent
  magnitude which actually is \sext's \texttt{MAG\_AUTO}.
\item $\log_{10}(Isophotal\  Area)-V$, where $Isophotal\ Area$ is the
area in pixels of each object above the threshold.
\item  $\log_{10}(Isophotal\ Area)-\log(\fwhm)$.
\item  $V-(V(r\leq1'')-V)$,  where V($r\leq1''$)  is   the  magnitude
  measured in an aperture of radius=1''.
\end{itemize}

In all these diagrams, stars populate a narrow and well defined region
while galaxies are more spread throughout the plane. This clear
segregation decreases for smaller and fainter objects for which stars
and galaxies populate similar regions in the diagrams. Therefore,
these diagrams are not really useful for faint ($V\gtrsim20$) and
small objects ($log(Area(pixels))\lesssim2$).

Displaying these kinds of  diagrams for each cluster  field it was easy
to detect outliers that most probably represented objects with a wrong
classification as  well  as remaining  spurious  detections. The (very
few)  suspicious  objects were    then   visually checked   and  their
classification changed if needed.

At the end of this process we computed the degree of misclassification
of  the original \sext catalogs.  It  was found that the fraction of
misclassification was minimal, less than  $1\%$ of misclassified stars
and $<0.6\%$ of misclassified galaxies, up to  $V\sim 22$.  Therefore,
we  consider that the  remaining  misclassifications in this range  of
magnitude  should be even smaller.  For   fainter objects the previous
diagrams  lose their utility since  stars and galaxies populate almost
the same region and even the visual check becomes useless.

The  published catalogs will   be  regularly updated  to  correct  for
possible new spurious objects or misclassifications found so users are
encouraged to use the latest version that will be  published in WINGS web
site:
\texttt{http://web.oapd.inaf.it/wings/}.

\section{The catalogs\label{sec:Catalogs}}

\subsection{Catalogs Description}

We have constructed three catalogs for each field containing
respectively galaxies, no saturated stars and objects of
unknown classification.  The structure of all of them is the same and
an example is shown in Table~\ref{tab:Catalogs}.

The parameters   stored    for each object   are   the  following  (in
parentheses we give the  name of the  output parameter from \sext\ used to
calculate them):
\begin{itemize}
\item  Equatorial     coordinates   (J2000.0)    of   the   barycenter
  (\texttt{X\_IMAGE,Y\_IMAGE}) and    of  the    peak    of   emission
  (\texttt{XPEAK\_IMAGE,YPEAK\_IMAGE}), taken from the V band image.
\item Isophotal area (\texttt{ISOAREA\_IMAGE}).
\item Kron radius (\texttt{KRON\_RADIUS}).
\item Full width at half maximum (\texttt{FWHM\_IMAGE}).
\item  Axis ratio  computed  as the ratio  of  the  \sext's parameters
  \texttt{B\_IMAGE} and \texttt{A\_IMAGE}.
\item   Position   angle with respect   to     the North and  measured
  counter-clockwise (\texttt{THETA\_IMAGE}).
\item \sext's \texttt{FLAG}
\item \sext's stellarity index (\texttt{CLASS\_STAR})
\item Concentration index  measured as the  ratio between the flux  in
  the central 9\%\footnote{Due to a bug found  in the modified version
    of \sext\ used to compute the  concentration index, the flux is not
    measured in the central 30\% of the pixels but in the central 9\%.} of the
  pixels and the total flux (\texttt{CONCENTRATION}).
\item  Contrast index measured  as the  ratio  between the flux in the
  brightest 30\% of the pixels and the total flux (\texttt{CONTRAST}).
\item Surface brightness of the brightest pixel (\texttt{MU\_MAX}).
\item And the following magnitudes in both bands:
  \begin{itemize}
  \item Total magnitudes: \sext's \texttt{MAG\_ISO},
    \texttt{MAG\_ISOCOR} and \texttt{MAG\_AUTO}
   \item Three  magnitudes at fixed  physical  apertures (at the target
    cluster's redshift): R=2\,kpc, 5\,kpc, 10\,kpc.
  \item  Three  magnitudes at    fixed  angular  apertures:  1\farcs6,
  2\farcs0, 2\farcs16\footnote{These  fixed apertures have been chosen
  for the multifiber spectroscopy of the WINGS fields. 1\farcs6
  is the projected diameter of the fibers in Autofib2@WHT while in the
  2dF@AAT the diameter varies radially in the field from 2\farcs16 in
  the center to 2\farcs0 in the edges. }
  \end{itemize}
\end{itemize}

The photometric catalogs are public only in  electronic format via CDS
and  at WINGS'   website \texttt{http://web.oapd.inaf.it/wings/}.  The
reduced images in both  bands will be made  available upon request to
the authors.

In the following  we give  indications  about some points to  be taken
into account when using the catalogs.
\begin{description}
\item[\textbf{Coordinates}] The astrometry of the images is discussed
  in Appendix~A.4 of \citetalias{Fasano2006}\footnote{The astrometry
    was done with respect to the USNO-A2 Catalogs which result in
    slight differences ($\lesssim 0\arcsec3$) with respect the
    USNO-B1.}. The overall quality is very satisfactory (see Appendix
  A.4 of Paper I) with uncertainties of the order of 0\farcs03 for stars and
  ranging from 0\farcs18 for WFI@ESO images to 0\farcs25 for WFC@INT
  ones.  For each object two positions have been computed: that of the
  barycenter of the light distribution and that of the peak of the
  emission.  For most objects, they are almost coincident.  However
  there are two situations in which it is better to use one and not
  the other.  First, for faint low surface brightness galaxies the
  barycenter position is preferable since the position of the peak is
  highly affected by the noise.  On the other hand and most
  importantly, for large extended galaxies (such as the brightest cluster
  galaxies) the coordinates of the peak emission are a better choice
  to locate the center of the object because the barycenter is
  strongly affected by the shape of the most external isophotes.
\item[\textbf{Shape   parameters}] Care must  be taken  when using the
  shape parameters especially   for objects in  crowded  fields and/or
  contaminated by close  companions.   The use  of  the \texttt{FLAGS}
  values can help to check this.
\item[\textbf{Total magnitudes}] \sext\ gives four different measures
  of total magnitude.  \texttt{MAG\_ISO} is the integrated light above
  the detection threshold.  \texttt{MAG\_ISOCOR} is an attempt to
  correct for the light lost in the wings of the objects making use of
  a Gaussian model approximation.  It was found that for B band
  magnitudes computed in dual mode \texttt{MAG\_ISOCOR} can give wrong
  values so we discourage its usage.  \texttt{MAG\_AUTO} is another
  attempt to guess the total magnitude of the galaxies using Kron
  apertures \citep{Kron1980,Infante1987,Bertin1996}.  However, several
  works \citep[e.g.][]{Franceschini1998} have shown the problems of
  this approximation, especially its dependence on the light profile
  of the object.  In Section~\ref{sec:PhotErrors} we will show that
  while for stellar and exponential profiles the \texttt{MAG\_AUTO}
  magnitude is a good approximation, for de Vaucouleurs profiles there
  is a systematic offset of $\sim0.2^m$\footnote{A detailed
    explanation of the Kron magnitude and its problems can be found in
    \citet{Graham2005}}.  Finally, \sext\ includes another total
  magnitude called \texttt{MAG\_BEST} magnitude.  This corresponds to
  \texttt{MAG\_AUTO} except in those cases in which the light of the
  object has a contamination by external sources higher than $10\%$,
  when \texttt{MAG\_ISOCOR} is used instead.  Several sources
  including the last \sext's Users Guide discourage the usage of
  \texttt{MAG\_BEST} in favor of \texttt{MAG\_AUTO}.  Our experience
  when constructing the catalogs has convinced us that 
    \texttt{MAG\_AUTO} gives the best approximation to the total
    magnitude of objects.
\item[\textbf{Saturated stars}] Since the aim of the catalogs is the
  study of clusters of galaxies, the saturated stars were discarded.
  This should be taken into account when using our catalog of stars.
  In the case of WFC@INT fields the incompleteness produced by
  saturation affects stars brighter than $V\sim17$, while for WFI@ESO
  fields this limits is $V\sim16$, the difference mainly from the
  different pixel size.
\end{description}

In addition to the information on the single objects found in each
field, in Appendix~\ref{sec:AddInfo} the reader can find a set of
tables with additional information about the target clusters,
peculiarities of the fields and the conditions of observation that
could be useful when working with our published data.

%
%
\subsection{Quality checks}

In this section we describe the internal photometric
quality\footnote{The comparison with external data (SDSS) was
    already done in \citetalias{Fasano2006}, section 5.4.} of the
data and the accuracy of the star/galaxy classification in our
catalogs.

\begin{figure}
  \centering
    \includegraphics[scale=0.45]{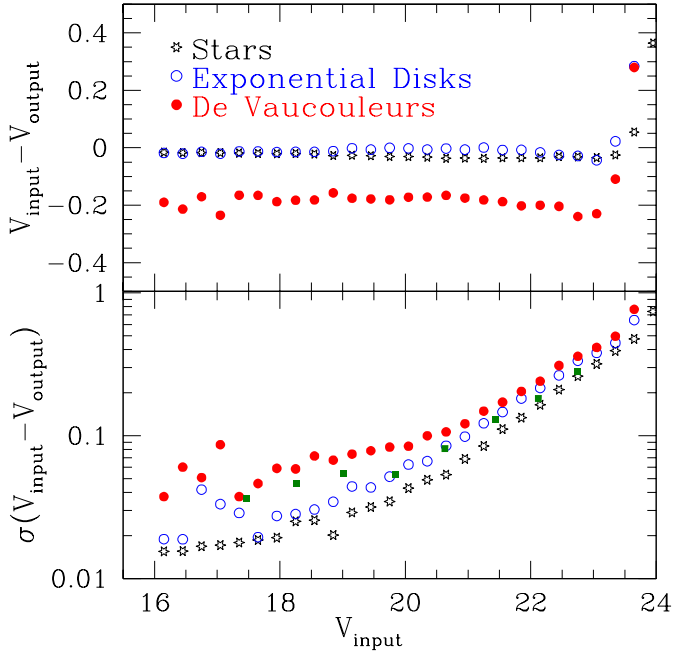}
    \caption{Photometric errors computed from simulations. Upper panel:
      Systematic offset of  the measured magnitude ($V_{output}$) with
      respect  to the input magnitude ($V_{input}$)  as  a function of
      the input magnitude. Lower panel: Dispersion of this
      difference. Green squares show the differences between the
      photometry of two fields (A780 and A970) observed with WFC@INT
      and WFI@ESO (same as lower fifth panel of Fig.11 in \citet{Fasano2006}.
      \label{fig:DifV_Sim}}
\end{figure}
\subsubsection{Photometric errors\label{sec:PhotErrors}}

When running without weighted images, \sext\ computes just the
photon-noise errors of the magnitudes.  For our deep images these
values are unrealistically small and for this reason we do not report
such errors in our catalogs.  In fact, the errors in the photometry
are mostly related to local background variations or to contamination
by the light of close companions.  To estimate these uncertainties we
performed simulations in which synthetic stars as well as galaxies
with exponential and de Vaucouleurs profiles were inserted in the
original images.  Fig.~\ref{fig:DifV_Sim} (upper panel) shows the
differences found between the input magnitude ($V_{input}$) and that
measured by \sext\ ($V_{output}$) as a function of the input magnitude
for the three types of objects used in the simulations: stars and the
two types of galaxy models.  Each point is the central biweight
estimator \citep{Beers1990} of the difference ($V_{input}-V_{output}$)
for objects in an interval of $\Delta(V_{input})=0.3^m$.  In the lower
panel we show the scale biweight estimator of $V_{input}-V_{output}$
in the same intervals.  Note the use of a logarithmic scale in
the ordinate axis.

It  is clear that for  de Vaucouleurs profiles,  \sext\ produces total
magnitudes that     are $\sim0.2^m$   fainter  than   real    ones, or
equivalently,  fluxes  that are $\sim20\%$ lower.    This is a known
problem of    \sext\ \citep{Franceschini1998,Benitez2004} although the
differences reported between the real and the measured magnitudes vary
from  work to work.   Contrary  to previous  works,  we find  that the
difference between input and output magnitudes remains almost constant
up to our detection limit ($V\sim23$).  On the other hand, stellar and
exponential profiles   show good   agreement  with  mean   differences
$\lesssim0.05^m$.

The simulations also give us a more realistic measure of the
uncertainties on the magnitudes (lower panel of
Fig.~\ref{fig:DifV_Sim}).  At fixed input magnitude, the largest
uncertainties are for galaxies with de Vaucouleurs profiles and the
lowest are for stars.  The trend with  input magnitude is similar for
the three types of objects with a steeper increase of the
uncertainties at magnitudes fainter than $V\sim20$. As a comparison
the figure also shows (green points) the differences in magnitudes
found when comparing the data from two fields (A780 and A970) observed
with our two instrumental set-ups (see Fig.~11 from
\citet{Fasano2006}). Therefore, the typical uncertainties in our
catalogs up to $V\sim 21.5$ is less than 0.1 mag. For magnitudes fainter
than $V=21.5$, it is possible to estimate the typical uncertainties
with the following expressions:
\begin{eqnarray*}
  \label{eq:PhotErr}
  \sigma_V(Star)    & = & 10^{0.309 V - 7.62}\\
  \sigma_V(Exp)     & = & 10^{0.271 V - 6.67}\\
  \sigma_V(deVauc)  & = & 10^{0.259 V - 6.34}
\end{eqnarray*}
%
%
\begin{figure}[htbp]
  \centering
  \includegraphics[scale=0.42]{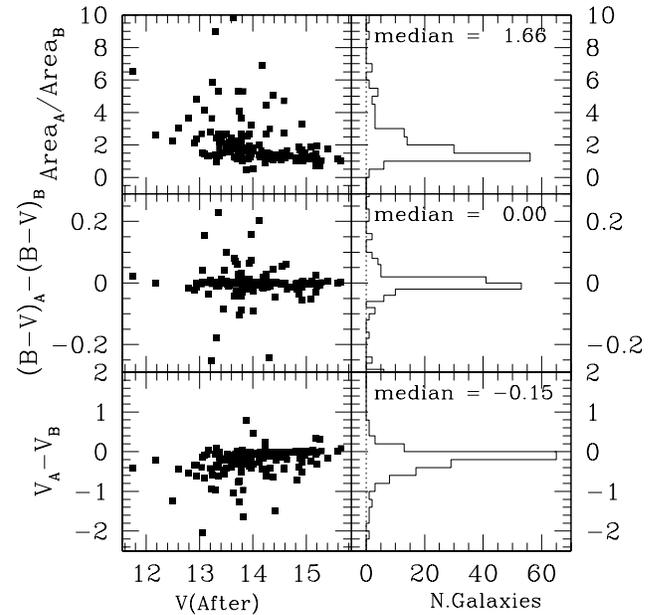}
  \caption{Comparison of the values of the total magnitude (lower
    panels), color index \bv (middle panels) and isophotal area (upper
    panels) measured by \sext\ of the modeled galaxies before and
    after separating them from the rest of the objects.  Left:
    Distribution of the differences of these parameters as a function
    of the total V magnitude.  Right: Distribution of the same
    differences. Index ``B'' denotes the values before applying our
    procedure and index ``A'' indicates the values after applying our
    procedure.}
  \label{fig:GalModelPhot}
\end{figure}
\begin{figure}
  \begin{center}
    \def\scl{0.45}
    \includegraphics[scale=\scl]{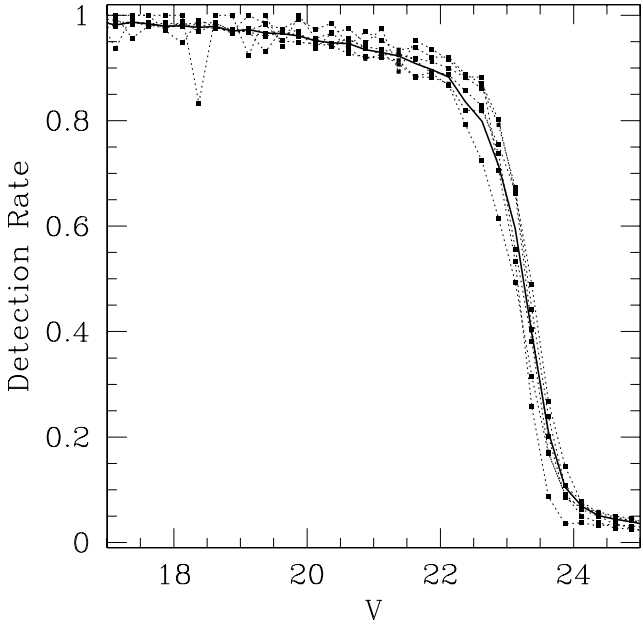}
    \caption{Average detection rate in each observing run computed
      from simulations. The stronger black  line is the detection rate
      averaged over the 77 fields.
      \label{fig:completeness}}
  \end{center}
\end{figure}
\subsubsection{Photometry of the modeled objects\label{sec:photmod}}

In this section we check the improvement in the photometry of the
objects modeled after applying the procedure described in
Appendix~\ref{sect:ap1} to the images.

Figure~\ref{fig:GalModelPhot} shows the effect of the model
subtraction on the output from \sext\ of some photometric parameters
of the modeled galaxies.  The large increase in the isophotal area
after applying our procedure clearly reflects the issue with the
segmentation of  large galaxies (see Fig.~\ref{fig:A193bcg}b).
This also affects the computation of the total magnitudes, as can be
seen in the lower panel of Fig.~\ref{fig:GalModelPhot}.  On the other
hand, the color index (B-V) shows smaller variations because of the
use of aperture magnitudes to compute it.

In addition to  the improvement  in  the photometry  of the subtracted
objects we also increased the detection rate of objects projected onto
the modeled halos. In the case of objects  projected onto the halos of
the  BCGs we found  $\sim 16\%$ more objects  after the subtraction of
the   BCG.  This is  a  relevant result especially   in studies of the
spatial distribution of galaxies in the cluster.

%
%
\subsubsection{Completeness and goodness of the star/galaxy
  classification\label{sec:Completeness}}

The simulations described above were originally intended also to check
the detection rate (completeness) and the success rate of the star/galaxy
classification.  However, we realized that the second step could not be
done due to an unexpected issue.  We found that commonly used programs
to simulated objects (stars and galaxies) do not work well at faint
magnitudes.  The result is that when running \sext, simulated stars
have a lower probability of being well classified than real stars (see
Appendix~\ref{appendix2}).  For that reason, simulations were used
only to estimate the detection rate in each field.

\begin{figure}
  \begin{center}
    \includegraphics[scale=0.45]{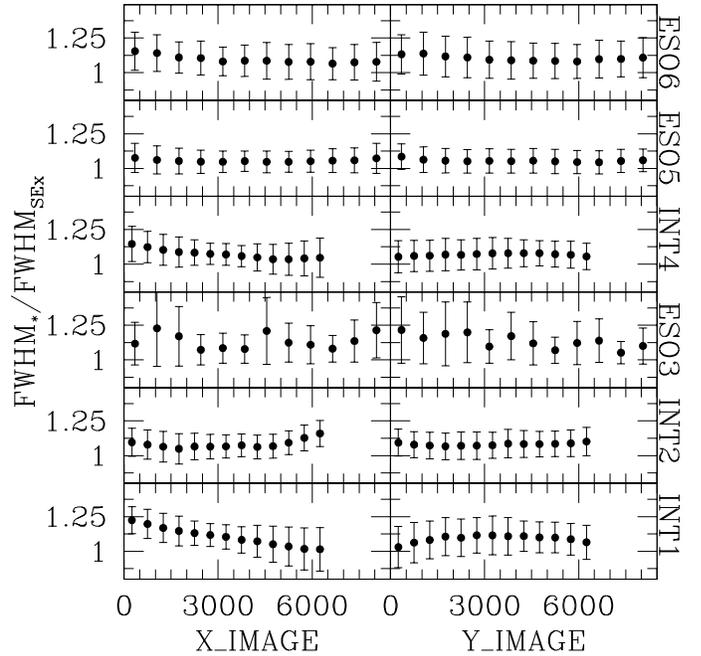}
    \caption{Spatial variation of the stellar FWHM in the different
      runs of WINGS.   Each point is   the central biweight  estimator
      (CBE,\citet{Beers1990}) of the ratio between the FWHM ($FWHM_*$)
      and the FWHM used as input in \sext\ ($FWHM_{SEx}$) for stars in
      bins  of  500 pixels (WFC@INT) or   700 pixels (WFI@ESO) along
      both axis.  The errors have  been computed as the scale biweight
      estimator (SBE).
      \label{fig:StellarFWHM}}
  \end{center}
\end{figure}

Figure~\ref{fig:completeness}  shows the average  detection rate  run by
run (dotted curves) and  the total average detection rate  (continuous
line).  The curves are quite similar and  we can considered our global
catalogs $90\%$ complete for objects with $V\lesssim21.7$.  The $50\%$
detection level is reached at $V\sim23.2$.  In addition, we provide in
Table~\ref{tab:Detection} the V band magnitudes at which the detection
rate  drops to $90\%, 75\%$  and $50\%$ in each field   as well as the
surface brightness detection thresholds in the same band.

As an external check of the completeness computed with the simulations
we took the image of the center of A1795 taken from the Space
Telescope and compared the detection rates in that image with ours.
The results are presented in Appendix~\ref{sec:appendix3} and they are
in good agreement with those from the simulations.

Concerning the star/galaxy classification, for objects with
\mbox{$V\lesssim21$} the difference between light profiles of stars
and galaxies is large enough to allow \sext\ to distinguish them
easily and, moreover, the diagrams like those in Fig.~\ref{fig:Vmumax}
can be used to reclassify the cases where \sext\ fails. Therefore, for
these bright objects the number of misclassifications can be
considered negligible.

\begin{figure}[t]
  \begin{center}
    \def\scl{0.45}
    \includegraphics[scale=\scl]{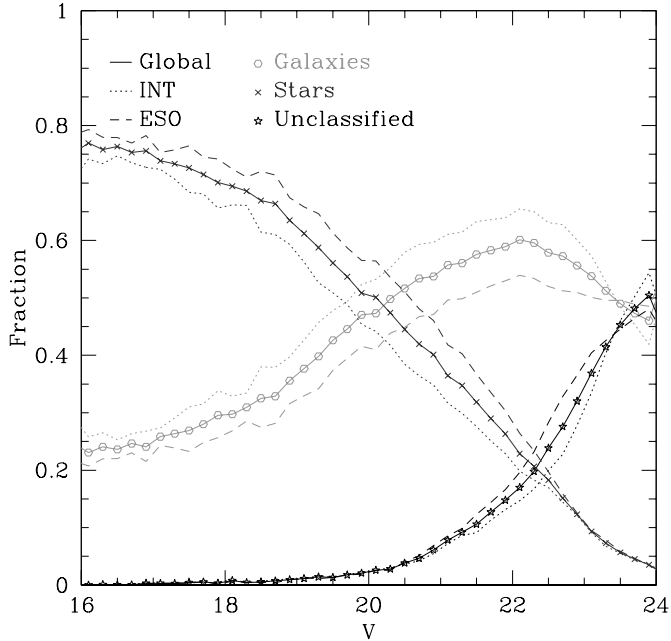}
    \caption{Relative numbers of the three types of objects:  galaxies,
      stars  and unknown objects. WFI@ESO fields show a higher fraction
      of stars because some of them point closer to the direction of
      the galactic center.
      \label{fig:fraction}}
  \end{center}
\end{figure}

\begin{figure}[t]
  \begin{center}
    \def\scl{0.45}
    \includegraphics[scale=\scl]{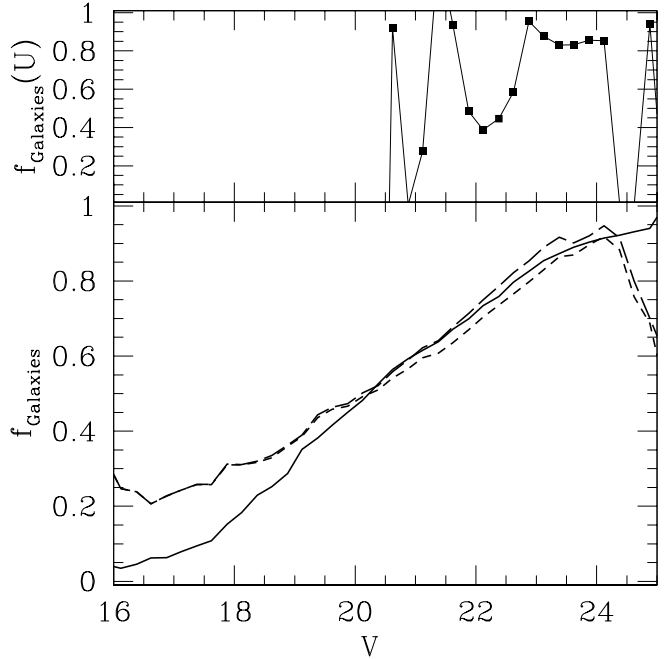}
    \caption{Lower panel: Average fraction of galaxies as a function of the V
      magnitude.   Continuous  line: Fraction  computed   assuming the
      stellar counts from the models of Besan\c{c}on and galaxy counts
      from \cite{Berta2006}.  Short dashed line: Data  from off-center
      regions ignoring  the unknown objects.    Long dashed line: Data
      from off-center   regions  including   the unknown   objects  as
      galaxies. Upper panel: Minimum fraction of galaxies in the
      unknown objects.
      \label{fig:ratioGalStar}}
  \end{center}
\end{figure}

The problem arises for fainter objects whose profiles become more
similar and small variations in the local conditions (e.g.  small
variations in the background) can make \sext\ fail to perform a good
classification.  Moreover, the number of objects is so large that
interactive cleaning is unpractical and even the visual check cannot
help.  This is worsened by the fact that the point spread function
actually varies throughout the field (Fig.~\ref{fig:StellarFWHM}).
For these reasons, a statistical approach appears to be the best
option to check the reliability of the star/galaxy classification at
faint magnitudes.  As stated above, the simulations were found to be
unreliable at faint magnitudes so we opted for another way to estimate
the goodness of our star/galaxy classification as well as to guess the
composition of the ``unknown'' class of objects.  First of all, we can
compare the fraction of objects in each group (stars/galaxies/unknown)
as a function of the V magnitude (see Fig.\ref{fig:fraction}). As
expected, the fraction of unknown objects increases at fainter
magnitudes. This increase rises more steeply for $V\gtrsim 22$, i.e.
when the completeness starts to fall rapidly.  A interesting point
is that this rise corresponds to a change in the trend of the
fraction of galaxies, pointing to the fact that most unknown objects
should be galaxies. Another way to see this is by relying upon external
sources for galaxy and star counts.  For the galaxies we took data
from the ESO-Spitzer Imaging extragalactic Survey \citep[ESIS,
][]{Berta2006} while for the stars we used the models of distribution
of stars in the Galaxy of the Observatory of
Besan\c{c}on\footnote{http://bison.obs-besancon.fr/modele/
}~\citep{Robin2003}.  Of course, this check is relevant only in the
faintest part of the magnitude distribution where counts from stars
and background galaxies dominate those of the target cluster.  With
both counts we computed the fraction of galaxies with respect to the
total number of objects as a function of the apparent magnitude
(continuous line in Fig.~\ref{fig:ratioGalStar}). This is compared
with the same fraction measured from WINGS fields (dashed lines)
although measured in off-center regions to avoid the high contribution
of galaxies from the center of the clusters. If the ``unknown'' objects
are ignored (short-dashed line in the figure) the fraction of galaxies
observed in our fields becomes smaller than that observed in the 
ESIS+Besan\c{c}on sample for $V\gtrsim 20.5$. This is an unexpected
result since our fields include the contribution of the cluster. To check
the influence of the classification of galaxies as ``unknown''
objects, we assumed that all the ``unknown'' objects were galaxies and
recomputed the fraction (long-dash line).  This produces an
overestimation of the fraction of faint galaxies. It is possible to
compute the minimum fraction of galaxies in the ``unknown'' sample
which makes the WINGS galaxy fraction equal that of the background
counts of galaxies and stars.  This is shown in the upper panel of
Fig.~\ref{fig:ratioGalStar} and for $V>23$ this fraction is greater
than $80\%$.

The agreement  between the fraction  of galaxies observed and expected
from external sources reassures  us  that even  at very faint  magnitudes
($V\sim24$)   the  star/galaxy classification  is   not  a mere random
assignment  from which one would  expected  half  stars and  half
galaxies. Since the difference in  the fraction of galaxies
is always $<  10\%$ between our  data and  those  from the external
sources     we    can  assume   that     statistically the star/galaxy
classification holds quite well up to $V\sim24$.

\section{Summary\label{sec:Summary}}

In this  paper   we present  the  first data  release  from  the WINGS
project.    We have produced photometric   catalogs of all the objects
found in 77 fields centered on nearby clusters.  The objects have been
classified as galaxies, stars and  objects of unknown type.  For  each
object we give positions, geometrical  parameters and different  kinds
of total and aperture magnitudes in both observed bands, B and V.

Our  catalogs are 90\%   complete at $V\sim  21.7$  and 50\% at $V\sim
23.2$ although  these  values vary from  field  to field, especially
depending on the total  exposure time.  Our star/galaxy classification
relies on the   \texttt{CLASS\_STAR} parameter of  \sext\ but  we have
performed a visual checking  of the brightest objects  ($V\lesssim22$)
that  makes the  misclassifications  negligible.   For   the faintest
objects we tried to use simulations  but we found  that these were not
reliable  in the appropiate   range of brightness ($V>22$).   However,
when we compare the  fraction of galaxies found  in our catalogs  with
those from external sources (ESIS' deep galaxy  counts and stars counts
from Observatory  of Besan\c{c}on's Galaxy  Model),  we find that both
are compatible  within $\pm 10\%$.   This result makes us confident in
our star/galaxy  classification  even  at faint magnitudes,  at  least
statistically. Again, the   reliability of  the  classification varies
from field to field, depending  on the seeing  and also on the spatial
variation of the PSF.

A notable feature of these catalogs is that extended galaxies have
been treated in a special manner which has allowed us to improve their
photometry, with several cases in which the total magnitudes were
initially underestimated by more than one magnitude.  This is a
critical issue for our fields, since our clusters are often centered
on very bright, extended galaxies, which are themselves important
targets of our study.  We have been able not only to improve the
photometry of these galaxies but also to detect $\sim 16\%$ more
objects around the BCG relative to the case in which the galaxy is
not subtracted.

In addition to the catalogs we also release the reduced images in both
bands (B,V)  and the preprocessed images  with the large halos removed
as well as the images with only the largest galaxies.

The scientific   analysis  of these  data will   be  published in  the
subsequent  papers of this series.

\begin{acknowledgements}
  The authors acknowledge the referee for the useful comments that
  helped to improve the initial manuscript.
 
  J. Varela  acknowledges a   post-doc fellowship FIRB/MIUR  from  the
  Ministero dell'Istruzione,  dell'Università  e della Ricerca (Italy)
  and a ``Juan de    la Cierva'' post-doc  fellowship granted   by the
  Spanish   ``Ministerio de   Educaci\'on y Ciencia''.   MM acknowledges
  support form the Spanish Ministerio de Educaci\'on y Ciencia, grants
  AYA2002-01241, AYA2005-07789 and AYA2006-14056.
   
  The 2.5m  ``Isaac   Newton''  Telescope  located  at Roque   de  los
  Muchachos (La Palma, Spain) is operated by the Isaac Newton Group of
  Telescopes on behalf of  the Particle Physics and Astronomy Research
  Council (PPARC) of the United Kingdom. The observations at the INT
  telescope were done under ``International Time Allocation''.
  
  The  2.2m Telescope  located  at La   Silla  (Chile) is   on loan to
  European Southern  Observatory    from the  Max  Planck  Gesellshaft
  (Germany).

  This  research has made use of  the NASA/IPAC Extragalactic Database
  (NED) which is operated by the Jet Propulsion Laboratory, California
  Institute of    Technology,   under  contract   with   the  National
  Aeronautics and Space Administration.
  
  IRAF (Image   Reduction and  Analysis   Facility)  is   written  and
  supported   by the IRAF programming  group   at the National Optical
  Astronomy Observatories (NOAO) in Tucson, Arizona.  NOAO is operated
  by the Association of Universities for Research in Astronomy (AURA),
  Inc.  under   cooperative   agreement with   the   National  Science
  Foundation.

\end{acknowledgements}

\appendix

\section{Full description of the extended halo removal\label{sect:ap1}}

In this appendix we give a detailed description of the whole procedure
followed to remove the objects with extended halos.

This is the  outline of the whole  process, while the single steps are
explained in some detail thereafter:

\begin{enumerate}
\item A background map is computed using the original images.
\item The extended objects   (galaxies and stars) to  be
  removed are selected interactively.
\item\label{point1} For each selected object, the following steps are performed:
  \begin{enumerate}

  \item The objects projected onto it are masked\footnote{The
  masking procedure consists of creating an auxiliary image or 
  \textit{mask} that indicates to the fitting program which pixels
  avoid to be in the fitting procedure.  This can be done, for
  example, by setting to zero all the pixels of the mask except those
  whose position correspond to pixels in the original image that
  should not be used in the fit.} with an interactive
  procedure.\label{mask}

  \item Its isophotes are fitted with the IRAF task \texttt{ellipse}.
  \item The resulting elliptical isophotes are then used as the input
    of the task \texttt{bmodel} to construct a model of the extended
    object.
  \item The model is subtracted from the original image.
  \item In the case of galaxies, when the subtraction leaves residuals
    due to   the structure of  the galaxy,  these are removed manually
    with \texttt{imedit}.
  \end{enumerate}
\item After  removing all the  selected  objects from the  initial image a new
  background map is calculated.
\item A mask of the remaining objects  is constructed with \sext. This
  mask  is  then  used   in  a   second  iteration  to   improve   the
  interactively-made   masks (point~(\ref{mask}))   applied  during  the
  isophote fitting procedure.
\item  At  this   point, the      procedure can   be  repeated    from
  point~(\ref{point1}) until the subtraction is satisfactory.
\end{enumerate}

The   process is iterative   and could be  repeated   as many times as
necessary but we found  that one iteration was  enough to reach a
good photometric quality.

\begin{figure}[t]
  \begin{center}
    \def\scl{0.45}
    \includegraphics[scale=\scl]{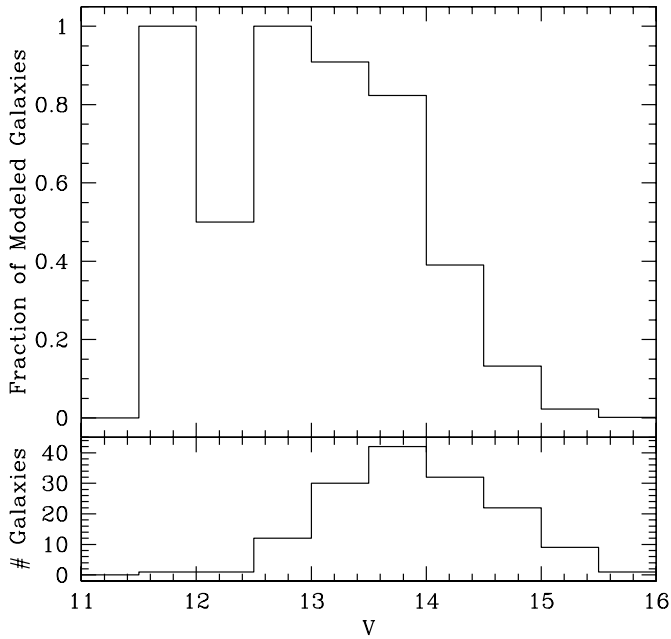}
    \caption{Upper panel: Fraction of galaxies that were
      modeled. Lower panel: Total number of galaxies that were modeled.
      \label{fig:VGalModel}}
  \end{center}
\end{figure}

Now we explain in more detail all the steps of the procedure:

\begin{description}
\item[\textbf{First  background computation}] Due to
  the variations found from chip to chip and to avoid the influence of
  the interchip regions  (which \sext\ treats as  if they were part of
  the actual image) the images were split into their original chips.
  
  After  that, an initial background  estimation was computed for each
  chip     using   \sext\      with    \texttt{BACK\_SIZE=256}     and
  \texttt{BACK\_FILTERSIZE=3} as input parameter values. To reduce the
  influence of  bright  pixels,   those  with intensities   above   an
  established threshold  were  replaced by  the  mode of the  pixel
  intensity of   the same chip,  which   can  be considered  as a first
  estimation of the  background  level.  The  threshold  was  set manually
  before starting the process seeking to not remove the bright part of
  the background.
  
  Once the background was obtained for each chip, the single
  background images were mosaiced to construct a global background
  map. This background map was then subtracted from the original
  image.
  
\item[\textbf{Selection of the objects to be removed}] The selection
  of the objects to be modeled and removed was done in a subjective
  way but following several guidelines.  First, the BCG was always
  modeled. Elliptical galaxies\footnote{Since the process is based on
    the construction of models of elliptical isophotes it is not
    suitable for spiral or irregular galaxies.}  highly blended and/or
  surrounded by small objects were also modeled.  The selection of the
  stars to be modeled was more field-dependent.  As the amount of work
  needed increased substantially with the number of objects to be
  modeled, in those fields with a high density of stars only the
  brightest stars were modeled.  Sometimes, a star was modeled to
  avoid the contamination from a close bright galaxy.
  
  Fig.~\ref{fig:VGalModel} shows the distribution in V of the galaxies
  that have been processed in this way. Up to $V\sim 14$ most of the
  galaxies have been modeled.
  
\item[\textbf{Fitting  and   modeling}]  To   reduce  the  computation
  time, the process of isophote fitting, modeling and model
  subtraction was done on small images of the selected objects
  extracted from the background-subtracted global image.

  
  The isophote fitting was  done using the IRAF task \texttt{ellipse}. 
  However,   before that, all   the  projected objects and problematic
  regions (e.g.  interchip  regions) were  masked.   Due to the  large
  number of objects that should be masked and  taking advantage of the
  iterative  process, this  first step  was  enough to make a
  rough mask. Below we will show how the improvement is achieved.

  
  A few differences were introduced when dealing with stars or
  galaxies.  For stars the ellipticity was fixed to zero and the
  fitting did not reach the innermost region (usually saturated and,
  therefore, not suitable for fitting isophotes).  Also, the center
  of the isophotes was not fixed because quite often the reflections of
  the stars were off-center with respect to the central regions.  For
  galaxies no restrictions were imposed.

  
  The output of \texttt{ellipse} was used as the input of the task
  \texttt{bmodel} which allows us to construct a two dimensional model
  of the object.  This model was subtracted from the original image
  (i.e.  before the first background subtraction).
  
  This step was done for all selected objects.

\item[\textbf{Second background estimation and object mask}] The outcome of the
  previous step was an image similar to the original one, in which
  large objects have been removed. This allowed us to obtain an
  improved background map, a better background subtracted image and a
  better detection of the small objects. In fact, using \sext\ with
  the background-subtracted images after removing  the models, it
  was possible to get a careful mask of all objects (except, of
  course, from those that were removed from the image).  This new
  mask would serve to refine the previous (manually-done) one.
  
\item[\textbf{Second fit and modeling}] The fitting process done in
  the first iteration was repeated using a refined mask for each
  object . This helped to get a better fit and, as a result, a better
  model.
  
  The main difference with respect to the first iteration is that
  after the subtraction of the model from the galaxies, sometimes some
  residuals remained in the central regions.  In such cases, these
  were manually edited using the IRAF task \texttt{imedit}.  When
  editing the images only clearly spurious residuals were replaced by
  pixels simulating the background signal adding a Gaussian noise
  whose sigma was computed from surrounding regions.  This has no
  effect on an object's photometry since the photometry of large
  galaxies is done with another image, avoiding the spurious detection
  of the residuals.
  
\item[\textbf{Construction of the  final images}] In most fields, only
  two iterations were enough to achieve a satisfactory result. After
  modeling all the selected extended objects and subtracting them from
  the initial image,  the background was recomputed again and finally
  subtracted from the model-subtracted image.  The output of this
  procedure was a background-subtracted image without the extended
  modeled objects. 

 The final step was to produce the complementary image containing only
 the modeled galaxies.  To  be consistent in   the photometry of   the
 extended galaxies we do not use the models. We constructed a new image
 containing the original pixels (background subtracted) of the removed
 galaxies. However,  to minimize the effect  of the projected objects,
 the intensities of   the pixels in which  these   fell were  in  fact
 replaced by the  intensities of the  models in the same pixels. In
 practice, this  was  done constructing  a   new mask of the   objects
 contained in  the background-subtracted  image without large  objects
 using \sext.  Then, the    final image was computed following   these
 criteria for its pixels:
  \begin{itemize}
  \item If the  pixel did not belong to  an  object (value  in the mask
    equal  to  zero)  then the  background-subtracted image  value was
    kept that corresponded to the original pixel of the galaxy.
  \item If  the pixel indeed belonged to  an object (value in the mask
    greater than zero)  then the value in the  model was taken instead
    of the value in the original image.
  \item If the pixel fell in an interchip region then the value from
    the model was used if available. In this way, we were able to
    improve the photometry of large galaxies with large regions lost
    in the interchip regions.
  \end{itemize}
  
  Of course, for construction, the image of the large galaxies also
  lacks the bright stars.

\end{description}

This procedure  was done in  both bands (V,B) removing,  of course,
the same objects.

%
%

\begin{figure}
  \centering
    \includegraphics[scale=0.45]{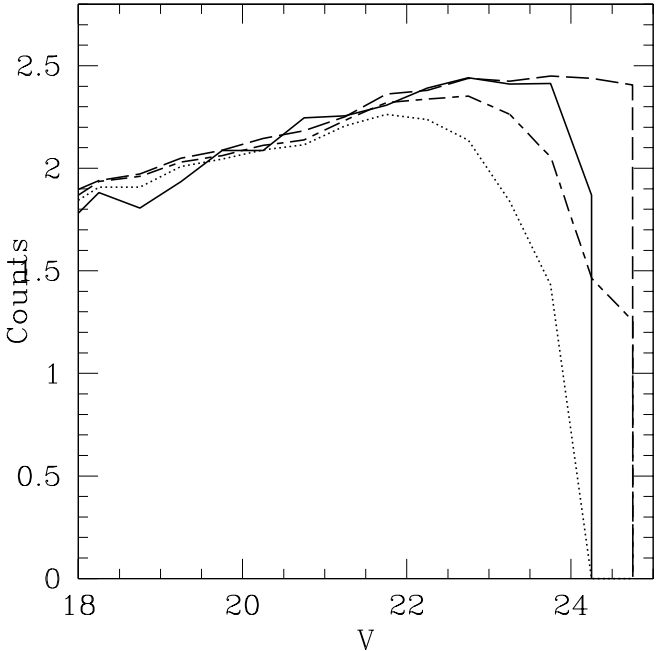}
    \caption{Comparison of the real counts (continuous line) with that
      coming from the simulations. Long dashes line: Input
      distribution of the simulated stars taken from Besan\c{c}on
      models. Dotted line: Simulated stars classified as
      stars. Long-short dashed line: Simulated stars detected.
      \label{fig:StarsRealSim}}
\end{figure}

\section{Image simulations\label{appendix2}}

One of the methods most commonly used to check the reliability of a
procedure of star/galaxy classification is to test the procedure using
images in which synthetic objects have been added. Of course, this
method relies upon the idea that the artificial objects are similar
enough (from the point of view of the classifying program) to the real
objects so that one can extrapolate the results obtained from the
simulations (from which the input and the output are known) to the
real objects.

We followed this method using the tasks of the IRAF's package
\texttt{artdata}.  We made catalogs of  galaxies and stars to build
the artificial images that were added to the real images.  In this way
we could test also the effects of the variations of the background and
other real conditions that are difficult or impossible to simulate.
The effects of crowding coming from adding more objects to real images
were measured to produce less than $~10\%$ of the lost objects in the
most crowded image.  Then we proceeded with these images as with the
real ones.  The resulting parameters from \sext\ were compared between
real and synthetic objects, showing that they were at first sight quite
similar, giving us confidence that the extrapolation from simulations
to real objects could be done.  However, the comparison of the final
counts from the real and the simulated objects made us distrust the
simulations, especially that of the faint stars.  The reason for this
conclusion is illustrated in  Fig.~\ref{fig:StarsRealSim}.
This figure shows the star counts from the original catalogs of WINGS
(continuous line) and from the simulations.  For simulated stars three
lines have been plotted.  The long dashed one represents the input
catalog and the numbers are taken from the models of the Galaxy from
Besan\c{c}on.  The long-short dashed line represents the counts of
detected simulated stars.  Finally, the dotted line shows the counts
of the simulated stars classified as stars, which should be the counts
to be compared with the data from WINGS' stars catalog (continuous
line).  From this figure it can be seen that, at the faintest
magnitudes, it was easier for \sext\ to distinguish a real star than a
simulated one, violating the initial premise that simulated objects
are similar to real ones.

In the  construction  of the simulated  stars we  took bright  but not
saturated stars to have  a well sampled   point spread function  (PSF)
even  at the  wings.   So, one   expects simulated stars  to   be more
concentrated  than real ones and then  easier to be classified as stars
by \sext, which is the opposite of what was  found.  The doubts became
greater when  simulations were done  using two different PSFs, one for
bright stars computed  from a bright start and  another  one for faint
stars computed from a fainter star. In this case, the fraction of faint
simulated stars identified as stars by  \sext\  increased.   Therefore, the
simulations  turn out to be too dependent on the input.

As this seemed to be a problem of the  \texttt{mkobjects} task we tried
with the \texttt{addstars}  feature of the \texttt{daophot} package to
create  the  synthetic stars. Since this   is a package  made to study
stars we expected a better treatment of the simulations.  However, the
results were similar.

The   origin of  the  difficulties   is  not  clear. Probably, small
variations of  the  local conditions where  objects  are added produce
large effects in faint simulated objects.

All  these  results convinced  us  not  to  use   the results  of  the
simulations to  measure       the reliability  of      our star/galaxy
classification   at faint magnitudes.  However,    we did not find such
problematic       behavior         at         bright       magnitudes.
Fig.~\ref{fig:StarsRealSim}  shows that the  problems of  detection of
simulated stars (before any classification as galaxy or star) start above
$V\sim22$   when       the      detection     rate      drops     (see
Fig.~\ref{fig:completeness}). For   these reason  we decided  to still
rely on the simulations  to estimate the   photometric errors and  the
detection rates since, unfortunately, there is no better procedure to
estimate such quantities.

%

\section{Comparison with WFPC2@HST image\label{sec:appendix3}}

\begin{figure}
  \centering
    \includegraphics[scale=0.45]{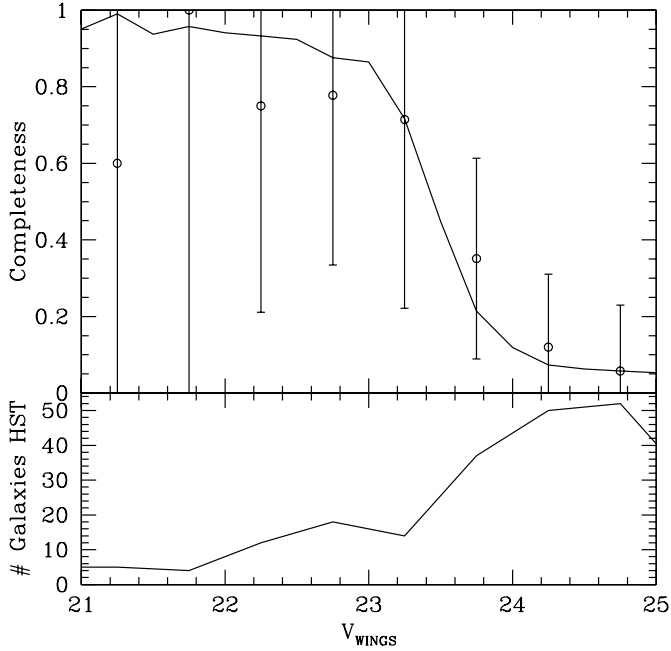}
    \caption{Comparison of the completeness of the WINGS catalog of
      the field of A1795. Upper panel: The continuous line shows the
      completeness computed from simulations while the dots with the
      error bars show the completeness when comparing from HST data.
      The error bars are constructing using $1\sigma$ Poisson
      errors. Lower panel: Number of objects detected in the HST
      image. Since the area is quite small the total numbers are also
      small producing the large uncertainties in the computation of the
      completeness.
      \label{fig:hst_comp}}
\end{figure}

To check the completeness estimations done with simulations we
performed a comparison using data from the Hubble Space Telescope. We
downloaded the images of the BCG of A1795 taken with the WFPC2 from
the HST archive{\footnote{Based on observations made with the NASA/ESA
    Hubble Space Telescope, obtained from the data archive at the
    Space Telescope Science Institute. STScI is operated by the
    Association of Universities for Research in Astronomy, Inc. under
    NASA contract NAS 5-26555.}}. We chose the F555W filter which was
the one that best matches the V band images that we used to construct
our catalogs.  From the mosaic of the WFPC2 we removed the PC chip in
which the BCG was located since this produced problems for \sext\ 
and we were interested in knowing the completeness at faint
magnitudes. After that, \sext\ was run on the image and the
resulting catalog was matched and compared with the WINGS catalog. We
performed the matching against the global WINGS catalogs, i.e.
including stars, galaxies and objects of unknown classification.
Fig.~\ref{fig:hst_comp} (upper panel) shows the completeness computed
for the A1795 field using the simulations (continuous line) and the
completeness compared with the HST data.  We also include  the errors
in the computation of these data since the area is quite small and
therefore the number of detected objects (lower panel) is also quite
low.

Although the HST image is much sharper than our ground-based image, it
is not much deeper and it also shows problems of completeness in the
range of comparison (lower panel of Fig.~\ref{fig:hst_comp}). This and
other issues in the matching procedure (such as pairs of objects that are
not resolved in the WINGS image) introduce uncertainties in the
comparison so this should be considered as a complementary check to
the completeness computed using the simulations. 

\section{Additional information\label{sec:AddInfo}}

In addition to the data of the single objects found in each field we
include several tables containing information about observational
features and peculiarities of either the single clusters or the fields
on which they are projected.

Position, redshift, Abell richness, Bautz-Morgan type, X-ray
luminosity and galactic extinction of the whole WINGS sample can be
found in Table 5 of \citetalias{Fasano2006}.

In Table~\ref{tab:UsefulData} we  summarize  the conversion  factors
used for  this work from  CCD related  units (pixels) to  angular unit
(arcsecs) and from these to linear units (kpc) at  the redshift of the
target   cluster using a  cosmological   model with parameters \mbox{$H_0=75$
km$\,$s$^{-1}\,$Mpc$^{-1}$},\ $\Omega_M=0.3$  and  $ \Omega_\Lambda=0.7$.    We have  also
included the effective total area of each image which is the real area
used to make the catalogs which is slightly smaller than the total field
of view   (for WFC@INT images, the effective    area is $\sim90\%$ of the
total field  of  view while  for WFI@ESO  this  value  is $\sim95\%$). 
We report the angular sizes of the  apertures of $R=2 $Kpc, 5 Kpc
and 10 Kpc used to construct our catalogs.

In Table~\ref{tab:Detection} we list the detection limits in surface
brightness, or surface brightness thresholds, ($\mu_V(Threshold)$) as
well as the V band magnitude at which the detection rate goes down to
$90\%, 75\%$ and $50\%$.  These last values are average values
obtained from the simulations.

Table~\ref{tab:BCG} reports the positions of the brightest cluster
galaxies.  In most cases we preferred the coordinates of the peak of
the emission instead of the coordinates of the barycenter because the
latter are more affected by irregularities in the outer isophotes.

Finally, Table~\ref{tab:Notes} lists comments or issues about the
clusters and the fields which we find interesting or useful when
working with our catalogs.

\clearpage

{\small
\clearpage 
\begin{longtable}{lccccc@{\extracolsep{\fill}}cccccc}
\caption{Useful parameters of the WINGS' clusters sample\label{tab:UsefulData}}\\
\hline\hline
Cluster & 
Redshift  & 
DM & 
\multicolumn{3}{c}{Plate Scale} &
\multicolumn{2}{c}{Total Area} &
\multicolumn{3}{c}{R$_{Aper}$ in arcsec} \\
\cline{4-6}\cline{7-8}\cline{9-11}
  &
  &
  &
arcsec/pixel & 
kpc/arcsec & 
kpc/pixel & 
deg$^{2}$ & 
Mpc$^{2}$ &
 2 Kpc & 5 Kpc & 10 Kpc \\
\hline
\endfirsthead
\caption{Useful parameters of the WINGS' sample (continued).}\\
\hline\hline
Cluster & 
Redshift  & 
DM & 
\multicolumn{3}{c}{Plate Scale} &
\multicolumn{2}{c}{Total Area} &
\multicolumn{3}{c}{D$_{Aper}$ in arcsec} \\
\cline{4-6}\cline{7-8}\cline{9-11}
  &
  &
  &
arcsec/pixel & 
kpc/arcsec & 
kpc/pixel & 
deg$^{2}$ & 
Mpc$^{2}$ &
 4 Kpc & 10 Kpc & 20 Kpc \\
\hline
\endhead
\hline
\multicolumn{12}{r}{{Continued on next page}} \\
\endfoot

\hline
\endlastfoot
    A85   &    0.0551 &    36.79  &    0.333  &     0.996  & 0.3317   &      0.2825  &    3.632    &  4.23 & 10.57 & 21.15 \\
    A119  &    0.0442 &    36.30  &    0.333  &     0.811  & 0.2701   &      0.2825  &    2.409    &  4.91 & 12.28 & 24.55 \\
    A133  &    0.0566 &    36.85  &    0.238  &     1.022  & 0.2433   &      0.2839  &    3.845    &  3.69 &  9.22 & 18.43 \\
    A147  &    0.0447 &    36.32  &    0.333  &     0.820  & 0.2730   &      0.2825  &    2.460    &  4.92 & 12.31 & 24.62 \\
    A151  &    0.0533 &    36.71  &    0.333  &     0.966  & 0.3217   &      0.2825  &    3.417    &  4.12 & 10.30 & 20.59 \\
    A160  &    0.0447 &    36.32  &    0.333  &     0.820  & 0.2730   &      0.2825  &    2.460    &  4.93 & 12.33 & 24.65 \\
    A168  &    0.0450 &    36.34  &    0.333  &     0.825  & 0.2747   &      0.2736  &    2.412    &  4.89 & 12.23 & 24.47 \\
    A193  &    0.0486 &    36.50  &    0.333  &     0.886  & 0.2950   &      0.2805  &    2.853    &  4.56 & 11.41 & 22.82 \\
    A311  &    0.0661 &    37.28  &    0.333  &     1.184  & 0.3942   &      0.2825  &    5.131    &  3.40 &  8.50 & 16.99 \\
    A376  &    0.0484 &    36.49  &    0.333  &     0.883  & 0.2939   &      0.2825  &    2.852    &  4.50 & 11.24 & 22.49 \\
    A500  &    0.0670 &    37.32  &    0.238  &     1.199  & 0.2854   &      0.2910  &    5.423    &  3.34 &  8.34 & 16.68 \\
   A548b  &    0.0416 &    36.18  &    0.238  &     0.767  & 0.1825   &      0.2910  &    2.218    &  4.94 & 12.35 & 24.71 \\
    A602  &    0.0619 &    37.09  &    0.333  &     1.112  & 0.3704   &      0.2762  &    4.430    &  3.58 &  8.96 & 17.92 \\
    A671  &    0.0502 &    36.57  &    0.333  &     0.913  & 0.3041   &      0.2762  &    2.986    &  4.37 & 10.93 & 21.86 \\
    A754  &    0.0542 &    36.75  &    0.333  &     0.981  & 0.3268   &      0.2768  &    3.455    &  4.08 & 10.19 & 20.38 \\
    A780  &    0.0539 &    36.73  &    0.333  &     0.976  & 0.3251   &      0.2768  &    3.419    &  3.92 &  9.80 & 19.60 \\
   A957x  &    0.0436 &    36.27  &    0.333  &     0.801  & 0.2667   &      0.2762  &    2.297    &  4.87 & 12.17 & 24.35 \\
    A970  &    0.0587 &    36.95  &    0.333  &     1.058  & 0.3523   &      0.2762  &    4.007    &  3.73 &  9.33 & 18.66 \\
   A1069  &    0.0650 &    37.23  &    0.333  &     1.165  & 0.388    &      0.2762  &    4.860   &  3.58 &  8.95 & 17.90 \\
   A1291  &    0.0527 &    36.68  &    0.333  &     0.956  & 0.3183   &      0.2762  &    3.271   &  4.18 & 10.46 & 20.92 \\
  A1631a  &    0.0462 &    36.39  &    0.238  &     0.845  & 0.2012   &      0.2874  &    2.661   &  4.69 & 11.74 & 23.47 \\
   A1644  &    0.0473 &    36.44  &    0.238  &     0.864  & 0.2056   &      0.2874  &    2.780   &  4.61 & 11.53 & 23.06 \\
   A1668  &    0.0634 &    37.16  &    0.333  &     1.138  & 0.3789   &      0.2762  &    4.636   &  3.52 &  8.79 & 17.58 \\
   A1736  &    0.0458 &    36.37  &    0.238  &     0.838  & 0.1995   &      0.2847  &    2.594   &  4.74 & 11.85 & 23.71 \\
   A1795  &    0.0625 &    37.12  &    0.333  &     1.122  & 0.3737   &      0.2762  &    4.509   &  3.58 &  8.95 & 17.90 \\
   A1831  &    0.0615 &    37.07  &    0.333  &     1.106  & 0.3682   &      0.2762  &    4.376   &  3.63 &  9.09 & 18.17 \\
   A1983  &    0.0436 &    36.27  &    0.333  &     0.801  & 0.2667   &      0.2715  &    2.257   &  4.91 & 12.28 & 24.55 \\
   A1991  &    0.0587 &    36.95  &    0.333  &     1.058  & 0.3523   &      0.2762  &    4.007   &  3.79 &  9.47 & 18.93 \\
   A2107  &    0.0412 &    36.16  &    0.333  &     0.759  & 0.2528   &      0.2872  &    2.146   &  5.32 & 13.30 & 26.61 \\
   A2124  &    0.0656 &    37.26  &    0.333  &     1.176  & 0.3916   &      0.2762  &    4.950   &  3.41 &  8.53 & 17.07 \\
   A2149  &    0.0679 &    37.36  &    0.333  &     1.215  & 0.4044   &      0.2762  &    5.281   &  3.31 &  8.28 & 16.56 \\
   A2169  &    0.0586 &    36.94  &    0.333  &     1.056  & 0.3517   &      0.2825  &    4.084   &  3.83 &  9.58 & 19.15 \\
   A2256  &    0.0581 &    36.92  &    0.333  &     1.048  & 0.3489   &      0.2820  &    4.013   &  3.82 &  9.54 & 19.09 \\
   A2271  &    0.0576 &    36.90  &    0.333  &     1.039  & 0.3461   &      0.2768  &    3.874   &  3.80 &  9.50 & 19.00 \\
   A2382  &    0.0618 &    37.09  &    0.238  &     1.111  & 0.2643   &      0.2916  &    4.662   &  3.46 &  8.66 & 17.32 \\
   A2399  &    0.0579 &    36.91  &    0.238  &     1.044  & 0.2486   &      0.2843  &    4.018   &  3.81 &  9.53 & 19.06 \\
   A2415  &    0.0581 &    36.92  &    0.333  &     1.048  & 0.3489   &      0.2740  &    3.898   &  3.80 &  9.49 & 18.98 \\
   A2457  &    0.0594 &    36.98  &    0.333  &     1.070  & 0.3563   &      0.2825  &    4.190   &  3.76 &  9.39 & 18.78 \\
  A2572a  &    0.0403 &    36.13  &    0.333  &     0.745  & 0.2480   &      0.2825  &    2.031   &  5.19 & 12.98 & 25.97 \\
   A2589  &    0.0414 &    36.17  &    0.333  &     0.764  & 0.2543   &      0.2825  &    2.134   &  5.26 & 13.16 & 26.31 \\
   A2593  &    0.0413 &    36.17  &    0.333  &     0.762  & 0.2537   &      0.2825  &    2.125   &  5.08 & 12.70 & 25.40 \\
   A2622  &    0.0620 &    37.10  &    0.333  &     1.114  & 0.3710   &      0.2825  &    4.544   &  3.63 &  9.07 & 18.15 \\
   A2626  &    0.0553 &    36.80  &    0.333  &     1.000  & 0.3330   &      0.2740  &    3.551   &  3.92 &  9.80 & 19.60 \\
   A2657  &    0.0402 &    36.12  &    0.333  &     0.743  & 0.2475   &      0.2825  &    2.022   &  5.41 & 13.52 & 27.04 \\
   A2665  &    0.0556 &    36.81  &    0.333  &     1.005  & 0.3347   &      0.2825  &    3.699   &  3.94 &  9.85 & 19.70 \\
   A2717  &    0.0490 &    36.51  &    0.238  &     0.893  & 0.2125   &      0.2832  &    2.926   &  4.41 & 11.03 & 22.06 \\
   A2734  &    0.0625 &    37.12  &    0.238  &     1.123  & 0.2672   &      0.2832  &    4.626   &  3.57 &  8.92 & 17.84 \\
   A3128  &    0.0599 &    37.00  &    0.238  &     1.078  & 0.2567   &      0.2916  &    4.394   &  3.76 &  9.41 & 18.81 \\
   A3158  &    0.0597 &    36.99  &    0.238  &     1.075  & 0.2558   &      0.2916  &    4.367   &  3.76 &  9.41 & 18.81 \\
   A3164  &    0.0570 &    36.87  &    0.238  &     1.029  & 0.2449   &      0.2842  &    3.900   &  3.64 &  9.10 & 18.20 \\
   A3266  &    0.0589 &    36.96  &    0.238  &     1.061  & 0.2526   &      0.2916  &    4.257   &  4.05 & 10.14 & 20.27 \\
   A3376  &    0.0456 &    36.36  &    0.238  &     0.835  & 0.1987   &      0.2916  &    2.635   &  4.71 & 11.78 & 23.57 \\
   A3395  &    0.0506 &    36.59  &    0.238  &     0.920  & 0.2190   &      0.2910  &    3.192   &  4.42 & 11.05 & 22.10 \\
   A3490  &    0.0688 &    37.40  &    0.238  &     1.230  & 0.2927   &      0.2874  &    5.634   &  3.21 &  8.03 & 16.06 \\
   A3497  &    0.0677 &    37.35  &    0.238  &     1.211  & 0.2882   &      0.2874  &    5.463   &  3.65 &  9.13 & 18.26 \\
  A3528a  &    0.0535 &    36.72  &    0.238  &     0.970  & 0.2307   &      0.2803  &    3.414   &  4.13 & 10.31 & 20.63 \\
  A3528b  &    0.0535 &    36.72  &    0.238  &     0.970  & 0.2307   &      0.2847  &    3.468   &  4.13 & 10.31 & 20.63 \\
   A3530  &    0.0537 &    36.72  &    0.238  &     0.973  & 0.2315   &      0.2837  &    3.480   &  4.06 & 10.15 & 20.31 \\
   A3532  &    0.0554 &    36.80  &    0.238  &     1.002  & 0.2384   &      0.2890  &    3.758   &  3.99 &  9.97 & 19.93 \\
   A3556  &    0.0479 &    36.47  &    0.238  &     0.874  & 0.2081   &      0.2874  &    2.846   &  4.48 & 11.20 & 22.40 \\
   A3558  &    0.0480 &    36.47  &    0.238  &     0.876  & 0.2085   &      0.2874  &    2.857   &  4.59 & 11.48 & 22.97 \\
   A3560  &    0.0489 &    36.51  &    0.238  &     0.891  & 0.2121   &      0.2874  &    2.958   &  4.66 & 11.64 & 23.29 \\
   A3562  &    0.0490 &    36.51  &    0.238  &     0.893  & 0.2125   &      0.2828  &    2.922   &  4.38 & 10.95 & 21.90 \\
   A3667  &    0.0556 &    36.81  &    0.238  &     1.005  & 0.2392   &      0.2822  &    3.696   &  4.16 & 10.41 & 20.81 \\
   A3716  &    0.0462 &    36.39  &    0.238  &     0.845  & 0.2012   &      0.2839  &    2.629   &  4.87 & 12.17 & 24.35 \\
   A3809  &    0.0620 &    37.10  &    0.238  &     1.114  & 0.2652   &      0.2838  &    4.566   &  3.53 &  8.83 & 17.65 \\
   A3880  &    0.0584 &    36.93  &    0.238  &     1.053  & 0.2506   &      0.2832  &    4.069   &  3.89 &  9.72 & 19.44 \\
   A4059  &    0.0475 &    36.45  &    0.238  &     0.867  & 0.2064   &      0.2832  &    2.761   &  4.57 & 11.42 & 22.83 \\
 IIZW108  &    0.0493 &    36.53  &    0.333  &     0.899  & 0.2993   &      0.2825  &    2.958   &  4.54 & 11.35 & 22.70 \\
   MKW3s  &    0.0450 &    36.34  &    0.333  &     0.825  & 0.2747   &      0.2762  &    2.435   &  4.82 & 12.05 & 24.10 \\
  RX0058  &    0.0470 &    36.43  &    0.333  &     0.859  & 0.2860   &      0.2825  &    2.700   &  4.66 & 11.64 & 23.29 \\
  RX1022  &    0.0491 &    36.52  &    0.333  &     0.895  & 0.2979   &      0.2762  &    2.865   &  4.13 & 10.33 & 20.67 \\
  RX1740  &    0.0430 &    36.25  &    0.333  &     0.791  & 0.2633   &      0.2768  &    2.243   &  5.06 & 12.65 & 25.29 \\
   Z1261  &    0.0644 &    37.20  &    0.333  &     1.155  & 0.3846   &      0.2825  &    4.884   &  3.46 &  8.66 & 17.32 \\
   Z2844  &    0.0500 &    36.56  &    0.333  &     0.910  & 0.3030   &      0.2762  &    2.964   &  4.40 & 10.99 & 21.98 \\
   Z8338  &    0.0473 &    36.44  &    0.333  &     0.864  & 0.2877   &      0.2768  &    2.678   &  4.63 & 11.57 & 23.15 \\
   Z8852  &    0.0400 &    36.11  &    0.333  &     0.740  & 0.2463   &      0.2825  &    2.003   &  5.41 & 13.52 & 27.04 \\
\end{longtable}}
{\small
\begin{flushleft}
  DM: Distance modulus.\\
  Total Area: This is the effective area used to construct the
  catalogs. To compute it the  interchip regions have been
  subtracted.\\
  $D_{Aper}$ : Conversion of the physical apertures to angular sizes.
\end{flushleft}
}
\normalsize

%
%

\begin{table}[h]
\normalsize
\caption{Completeness limits and surface brightness detection limits.\label{tab:Detection}}
\centering
\begin{tabular}{lcccc||lcccc}
\hline\hline
Cluster & 
\multicolumn{3}{c}{V detection rate limits } &
\multicolumn{1}{c||}{$\mu_V(Threshold)$} &
\multicolumn{1}{l}{Cluster} & 
\multicolumn{3}{c}{V detection rate limits } &
\multicolumn{1}{c}{$\mu_V(Threshold)$} \\
\cline{2-4}\cline{7-9}
 &
 $\ge 90\%$ &
 $\ge 75\%$ &
 $\ge 50\%$ &
$mag/arcsec^2$  &
 &
 $\ge 90\%$ &
 $\ge 75\%$ &
 $\ge 50\%$ &
$mag/arcsec^2$  \\
\hline\hline
A 85  &  22.6 & 22.9 & 23.2     &    25.85 &   A2589  &  22.4 & 23.1 & 23.5    &    25.89\\
A119  &  22.6 & 22.9 & 23.2      &    25.92 &   A2593  &  22.4 & 23.1 & 23.4    &    25.87\\
A133  &  22.3 & 22.8 & 23.1      &    25.56 &   A2622  &  21.6 & 22.9 & 23.3    &    25.93\\
A147  &  22.4 & 23.0 & 23.3      &    25.89 &   A2626  &  21.9 & 22.8 & 23.2    &    25.98\\
A151  &  22.2 & 22.8 & 23.2      &    25.73 &   A2657  &  22.4 & 22.9 & 23.3    &    25.93\\
A160  &  22.5 & 23.0 & 23.3      &    25.85 &   A2665  &  22.6 & 23.1 & 23.4    &    25.88\\
A168  &  22.6 & 23.1 & 23.4      &    25.90 &   A2717  &  22.4 & 22.8 & 23.2    &    25.12\\
A193  &  22.7 & 23.1 & 23.4      &    25.86 &   A2734  &  22.5 & 23.0 & 23.3    &    25.52\\
A311  &  22.6 & 23.1 & 23.4      &    26.00 &   A3128  &  22.1 & 22.7 & 23.2    &    25.54\\
A376  &  21.4 & 22.7 & 23.2      &    25.89 &   A3158  &  21.4 & 23.0 & 23.4    &    25.61\\
A500  &  22.8 & 23.2 & 23.5      &    25.53 &   A3164  &  21.5 & 22.8 & 23.2    &    25.71\\
A548b  &  21.8 & 22.9 & 23.3     &    25.42 &   A3266  &  21.3 & 22.9 & 23.3    &    25.46\\
A602  &  22.0 & 22.7 & 23.1      &    25.69 &   A3376  &  20.9 & 22.7 & 23.2    &    25.39\\
A671  &  22.1 & 22.7 & 23.1      &    25.75 &   A3395  &  21.3 & 23.0 & 23.3    &    25.35\\
A754  &  20.8 & 22.6 & 23.0      &    25.60 &   A3490  &  21.2 & 22.6 & 23.1    &    25.38\\
A780  &  20.7 & 21.6 & 22.2      &    25.29 &   A3497  &  22.2 & 23.2 & 23.7    &    25.60\\
A957x  &  22.0 & 22.6 & 22.9     &    25.64 &   A3528a  &  22.6 & 23.1 & 23.4   &    25.40\\
A970  &  21.7 & 22.6 & 22.8      &    25.57 &   A3528b  &  22.3 & 22.7 & 23.1   &    25.01\\
A1069  &  22.1 & 22.7 & 23.1     &    25.68 &   A3530  &  22.0 & 22.9 & 23.3    &    25.40\\
A1291  &  22.2 & 22.7 & 23.1     &    25.74 &   A3532  &  21.2 & 23.1 & 23.5    &    25.51\\
A1631a  &  22.5 & 23.1 & 23.5    &    25.39 &   A3556  &  21.4 & 23.1 & 23.5    &    25.56\\
A1644  &  22.7 & 23.2 & 23.5     &    25.42 &   A3558  &  21.8 & 23.0 & 23.5    &    25.56\\
A1668  &  21.9 & 23.1 & 23.4     &    26.12 &   A3560  &  21.1 & 23.1 & 23.5    &    25.59\\
A1736  &  22.1 & 23.1 & 23.6     &    25.49 &   A3562  &  19.8 & 21.6 & 22.0    &    25.47\\
A1795  &  22.6 & 23.1 & 23.4     &    25.96 &   A3667  &  19.7 & 22.1 & 22.9    &    25.56\\
A1831  &  22.5 & 23.0 & 23.3     &    25.83 &   A3716  &  21.3 & 22.6 & 23.0    &    25.53\\
A1983  &  22.7 & 23.1 & 23.5     &    26.00 &   A3809  &  22.5 & 23.1 & 23.5    &    25.54\\
A1991  &  22.7 & 23.1 & 23.4     &    26.00 &   A3880  &  22.6 & 23.0 & 23.3    &    25.52\\
A2107  &  22.5 & 22.9 & 23.2     &    25.78 &   A4059  &  22.4 & 23.0 & 23.3    &    25.49\\
A2124  &  22.6 & 23.0 & 23.3     &    25.84 &   IIZW108  &  19.8 & 22.1 & 23.0  &    25.86\\
A2149  &  21.6 & 22.0 & 22.3     &    24.65 &   MKW3s  &  21.6 & 22.1 & 22.4    &    24.80\\
A2169  &  22.4 & 23.0 & 23.3     &    25.85 &   RX0058  &  22.5 & 23.1 & 23.5   &    26.03\\
A2256  &  20.7 & 22.6 & 23.1     &    25.77 &   RX1022  &  22.7 & 23.1 & 23.3   &    25.75\\
A2271  &  21.5 & 21.9 & 22.2     &    24.79 &   RX1740  &  21.1 & 22.8 & 23.3   &    26.01\\
A2382  &  22.2 & 22.8 & 23.2     &    25.50 &   Z1261  &  20.8 & 22.7 & 23.2    &    25.82\\
A2399  &  22.1 & 22.7 & 23.2     &    25.48 &   Z2844  &  22.6 & 22.9 & 23.3    &    25.70\\
A2415  &  22.1 & 22.8 & 23.2     &    25.75 &   Z8338  &  21.4 & 22.9 & 23.4    &    26.03\\
A2457  &  22.1 & 22.9 & 23.2     &    25.82 &   Z8852  &  21.7 & 22.7 & 23.2    &    25.83\\
A2572a  &  21.9 & 23.0 & 23.4    &    25.93 & & & & &\\
\hline
\end{tabular}
\end{table}

%
%

\begin{table}[h]
\normalsize
\caption{Coordinates of the emission peak of the brightest cluster  galaxies\label{tab:BCG}}
\centering
\begin{tabular}{lrr||lrr}
\hline\hline
Cluster & 
\multicolumn{2}{c}{Equat.Coordinates (J2000.0)} &
Cluster & 
\multicolumn{2}{c}{Equat.Coordinates (J2000.0)} \\
 &
 \multicolumn{1}{c}{$\alpha$} &
 \multicolumn{1}{c}{$\delta$} &
 &
 \multicolumn{1}{c}{$\alpha$} &
 \multicolumn{1}{c}{$\delta$} \\
\hline
\hline
A 85   &  0:41:50.5 &  -9:18:11.5  & A2589  & 23:23:57.5 &  16:46:38.3  \\
A119   &  0:56:16.1 &  -1:15:19.1  & A2593  & 23:24:20.1 &  14:38:49.8  \\
A133   &  1:02:41.7 & -21:52:55.5  & A2622  & 23:35:01.5 &  27:22:21.0  \\
A147   &  1:08:12.0 &   2:11:38.2  & A2626  & 23:36:30.5 &  21:08:47.4  \\
A151   &  1:08:51.1 & -15:24:23.1  & A2657  & 23:44:57.4 &   9:11:35.3  \\
A160   &  1:12:59.6 &  15:29:28.9  & A2665  & 23:50:50.6 &   6:08:58.9  \\
A168   &  1:14:57.6 &   0:25:51.2  & A2717  &  0:03:13.0 & -35:56:13.3  \\
A193   & 1:25:07.6  &   8:41:57.2  & A2734  &  0:11:21.6 & -28:51:15.6  \\
A311   &  2:09:28.4 &  19:46:36.2  & A3128  &  3:29:50.6 & -52:34:46.8  \\
A376   &  2:46:03.9 & 36:54:19.2   & A3158  &  3:42:53.0 & -53:37:52.6  \\
A500   &  4:38:52.5 & -22:06:39.0  & A3164  &  3:45:25.8 & -56:59:02.0  \\
A548b  &  5:45:29.6 & -25:55:56.8  & A3266  &  4:31:13.3 & -61:27:12.0  \\
A602   &  7:53:26.6 & 29:21:34.5   & A3376  &  6:00:41.1 & -40:02:40.4  \\
A671   &  8:28:31.7 & 30:25:53.1   & A3395  &  6:27:36.3 & -54:26:57.9  \\
A754   &  9:08:32.4 & -9:37:47.4   & A3490  & 11:45:20.2 & -34:25:59.4  \\  
A780   &  9:18:05.7 & -12:05:43.4  & A3497  & 11:59:46.3 & -31:31:41.6  \\ 
A957x  & 10:13:38.3 & -0:55:31.3   & A3528a & 12:54:41.0 & -29:13:39.5  \\ 
A970   & 10:17:25.7 & -10:41:20.2  & A3528b & 12:54:22.2 & -29:00:46.8  \\  
A1069  & 10:39:43.4 &  -8:41:12.4  & A3530  & 12:55:36.0 & -30:20:51.4  \\ 
A1291  & 11:32:23.2 &  55:58:03.0  & A3532  & 12:57:22.0 & -30:21:49.1  \\
A1631a & 12:52:52.6 & -15:24:47.8  & A3556  & 13:24:06.7 & -31:40:11.6  \\
A1644  & 12:57:11.6 & -17:24:34.0  & A3558  & 13:27:56.8 & -31:29:44.0  \\  
A1668  & 13:03:46.6 &  19:16:17.4  & A3560  & 13:31:53.5 & -33:14:03.1  \\ 
A1736  & 13:27:28.0 & -27:19:29.2  & A3562  & 13:33:34.7 & -31:40:20.5  \\ 
A1795  & 13:48:52.5 &  26:35:34.6  & A3667  & 20:12:27.3 & -56:49:36.4  \\  
A1831  & 13:59:15.1 &  27:58:34.5  & A3716  & 20:51:19.9 & -52:38:10.5  \\
A1983  & 14:52:55.3 &  16:42:10.6  & A3809  & 21:46:59.1 & -43:53:56.2 \\
A1991  & 14:54:31.5 &  18:38:32.9  & A3880  & 22:27:54.4 & -30:34:31.9 \\
A2107  & 15:39:39.0 &  21:46:58.0  & A4059  & 23:57:00.7 & -34:45:32.9 \\
A2124  & 15:44:59.0 &  36:06:33.9  & IIZW108 & 21:13:55.9 & 2:33:55.4  \\
A2149  & 16:01:28.1 &  53:56:50.4  & MKW3s  & 15:21:51.9 & 7:42:32.1   \\
A2169  & 16:13:58.1 &  49:11:22.4  & RX0058 &  0:58:22.6 & 26:51:59.0  \\
A2256  & 17:04:27.2 &  78:38:25.4  & RX1022 & 10:22:10.0 & 38:31:23.9  \\
A2271  & 17:18:16.7 &  78:01:06.2  & RX1740 & 17:40:32.1 & 35:38:46.1  \\
A2382  & 21:51:55.6 & -15:42:21.3  & Z1261  &  7:16:41.2 & 53:23:09.5  \\
A2399  & 21:57:01.7 &  -7:50:22.0  & Z2844  & 10:02:36.5 & 32:42:24.3  \\
A2415  & 22:05:38.6 &  -5:35:32.1  & Z8338  & 18:11:05.2 & 49:54:33.7  \\
A2457  & 22:35:40.8 &   1:29:05.9  & Z8852  & 23:10:22.4 & 7:34:50.6   \\
A2572a & 23:17:12.0 &  18:42:04.7  & & &\\
\hline
\end{tabular}
\end{table}

%
%
%
\begin{longtable}{ll}
\caption{Remarks about the individual fields.\label{tab:Notes}}\\
\hline\hline
Cluster &
\multicolumn{1}{c}{Notes}\\
\hline
\endfirsthead
\caption{Remarks about the single fields (continued).}.\\
\hline\hline
Cluster &
\multicolumn{1}{c}{Notes}\\
\hline
\endhead
\hline 
\multicolumn{2}{r}{{Continued on next page}} \\
\endfoot
\hline
\endlastfoot

A85 & \\
A119 & \\
A133 & Two very bright stars in $(\alpha,\delta)_{J2000.0}$=(15.6155\degr,-21.6094\degr) and (15.5469\degr,-21.6080\degr)\\
A147 & Very bright star in (16.9983\degr,1.9926\degr)\\
A151 & Background structure at $z\sim0.096$\\
A160 & Background group at $z\sim0.060$\\
A168 & Merger of a spiral dominated cluster and a cD dominated one\\
A193 & \\
A311 & \\
A376 & Strong spatial variation of the PSF along the declination direction\\
     & Bright star in (41.1317\degr,36.7629\degr)\\
A500 & \\
A548b & Interacting central pair of galaxies\\
A602 & Cluster with two similar dominant galaxies \\
A671 & \\
A754 & Two bright stars at (137.1675\degr,-9.3841\degr)  and (136.9504\degr,-9.8537\degr)  \\
A780 & Bright stars at (139.5243\degr,-12.3193\degr) and (139.3175\degr,-12.0079\degr) \\
     & BCG contaminated by two stars that are at $\sim~1\,$arcmin\\
     & High contamination from a star outside of the field of view.\\
A957x & Very bright star at (153.6394\degr,-1.1178\degr)\\
A970 & \\
A1069 & \\
A1291 & Bright star in (173.0842\degr,56.0965\degr)\\
A1631a & Contaminated by a foreground group at $z\sim0.0144$\\
        & The brightest galaxy in the center of the field is a foreground galaxy\\
A1644 & Merging between two clusters \\
      & Bright star close to the center (194.3967\degr,-17.3951\degr)\\
A1668 & \\
A1736 & Cluster without a clear center and with two dominant galaxies
quite far away one from the other ($\gtrsim 14' \gtrsim 700\,kpc$).\\
      & Possible merger ? \\
A1795 & Bright star at (207.0113\degr,26.5154\degr) \\
      & Bright star at (207.2009\degr,26.6144\degr) at only $\sim1\farcm67$ of the BCG \\
A1831 & \\
A1983 & Two dominant galaxies\\
A1991 & Bright star at (223.4967\degr,18.4998\degr)\\
A2107 & Only field observed using high-dithering technique\\
A2124 & \\
A2149 & Very bright star at (240.1033\degr,53.9671\degr)\\
A2169 & Bright spike near the edge of one chip from (4050,1) to (4050,$\sim100$)\\
A2256 & Field with a high star density \\
      & Bright star at (255.8912\degr,78.6300\degr)\\
A2271 & Bright star in (257.0958\degr,78.8285\degr)\\
A2382 & Bright star at (328.1596\degr,-15.8079\degr)\\
A2399 & The brightest galaxy in the field is indeed a foreground ($z=0.017$) spiral galaxy\\
A2415 & Bright star at (331.5222\degr,-5.3578\degr)\\
A2457 & Contaminated by a galactic nebula\\
A2572a & Central pair of galaxies in interaction \\
        & Bright star at (349.4753\degr,18.7029\degr)\\
A2589 & Background group at $z\sim0.17$\\
A2593 & Bright star at (351.0991\degr,14.4786\degr) \\
A2622 & \\
A2626 & Contaminated by A2625 ($z=0.0609$) \\
A2657 & Field contaminated by a galactic nebula\\
A2665 & Bright stars at (357.6523\degr,6.0191\degr)\\
A2717 & Bright star at (0.8530\degr,-35.9656\degr) close to the BCG ($\sim2\farcm9 $)\\
A2734 & \\
A3128 & Bright star at (53.0046\degr,-52.4743\degr)\\
A3158 & \\
A3164 & BCG lost in interchip region !!\\
A3266 & Bright star at (67.9736\degr,-61.3848\degr)\\
A3376 & Bright star at (90.1511\degr,-39.9517\degr)\\
A3395 & \\
A3490 & \\
A3497 & About half of the BCG falls in an interchip region\\
A3528a & Double system together with A3528b\\
       & Member of a group of clusters with A3528a, A3530 and A3532 \\
      & Bright stars at (193.5528\degr,-29.4042\degr) and (193.4720\degr,-29.3077\degr)\\
A3528b & Only 460\,s of total exposure time in V and 180\,s in B\\
      & Double system together with A3528a\\
      & Member of a group of clusters with A3528b, A3530 and A3532 \\
A3530 & Member of a group of clusters with A3528a, A3528b and A3532 \\
      & BCG in interaction  with a close galaxy \\
      & Bright star at (193.8856\degr,-30.0642\degr)\\
A3532 & Member of a group of clusters with A3528a, A3528b and A3530\\
A3556 & Cluster with two dominant galaxies \\
A3558 & Bright stars at (202.0540\degr,-31.5522\degr) and (201.6474\degr,-31.4513\degr)\\
A3560 & The central galaxy of the field which is also the brightest galaxy in the field is a foreground galaxy ($z=0.0124$)\\
      & Bright star at (203.0631\degr,-33.2489\degr)\\
A3562 & Total exposure times of 180\,s in each filter\\
      & Very bad seeing ($FWHM_*(V)\sim2\farcs38$)\\
A3667 & Bright star at (303.2418\degr,-56.8468\degr)\\
A3716 & Brightest galaxy in the field is a background galaxy ($z=0.0557$) \\
      & Two bright central galaxies maybe in interaction \\
      & Bright star at (312.8122\degr,-52.6243\degr) \\
A3809 & Bright star at (326.4170\degr,-44.1396\degr)\\
A3880 & Bright star at (336.9604\degr,-30.6026\degr) at $\sim 1\farcm82$ from the BCG\\
A4059 & \\
IIZW108 & BCG in clear interaction \\
        & Very bright star at (318.1888\degr,2.6429\degr) \\
MKW3s & Two bright stars at (230.2168\degr,7.6827\degr) and (230.5907\degr,7.8160\degr), respectively\\
RX 58 & BCG with two nuclei separated by $\sim 7''$ \\
       & In the catalog they are considered as two objects \\
       & Two bright stars (14.6187\degr,26.7859\degr) and (14.6623\degr,26.8584\degr) both closer than   5 arcmin to the BCG\\
RX1022 & Problem with photometry found, probably extinction by
clouds. Corrected using SDSS data.\\
RX1740 & Field with high density of stars.\\
       & Bright star at (265.3072\degr,35.7383\degr) and (265.4106\degr,35.3667\degr)\\
Z1261 & \\
Z2844 & Very bright spike in the border of one chip (150.9810\degr,32.6458\degr)\\
Z8338 & Field with high density of stars\\
Z8852 & Very bright star at (347.5279\degr,7.3640\degr) \\
\end{longtable}


\begin{thebibliography}{}
\bibitem[Abell(1957)]{Abell1957} Abell, G.~O.\ 1957, "The distribution of rich clusters of galaxies. A catalogue of 2712 rich clusters found on the National Geographic Society Palomar Observatory Sky Survey". Chicago: Univiersity of Chicago Press.  
\bibitem[Abell(1959)]{Abell1959} Abell, G.~O.\ 1959, Leaflet of  the Astronical Society of the Pacific, 8, 121  
\bibitem[Abraham {et~al.}(1994)]{Abraham1994} Abraham, R.~G., Valdes, F., Yee, {et~al.} S.\ 1994, \apj, 432, 75  
\bibitem[Abraham {et~al.}(1996)]{Abraham1996} Abraham, R.~G., van den Bergh,  S., Glazebrook, K., {et~al.} 1996, \apjs, 107, 1  
\bibitem[Bahcall {et~al.}(2003)]{Bahcall2003} Bahcall, N.~A., {et~al.}\  2003, \apjs, 148, 243  
\bibitem[Beers, Flynn \& Gebhardt(1990)]{Beers1990} Beers, T.~C., Flynn, K. \& Gebhardt, K.\ 1990, \aj, 100, 32  
\bibitem[Ben{\'{\i}}tez {et~al.}(2004)]{Benitez2004} Ben{\'{\i}}tez, N., et  al.\ 2004, \apjs, 150, 1   
\bibitem[Berta {et~al.}(2006)]{Berta2006} Berta, S.,  Rubele, S., Franceschini, A., {et~al.}\ 2006,  \aap, 451, 881
\bibitem[Bertin  \&  Arnouts(1996)]{Bertin1996} Bertin, E.~\& Arnouts, S.\ 1996, \aaps, 117, 393  
\bibitem[{Biviano  {et~al.}(1997)}]{Biviano1997}  Biviano, A., Katgert, P., Mazure, A., {et~al.} 1997, \aap , 321, 84  
\bibitem[Butcher \& Oemler(1978)]{Butcher1978a} Butcher, H.~\& Oemler, A.\ 1978, \apj, 219, 18  
\bibitem[Cava et al.(2008)]{Cava2008} Cava, A., et al.\ 2008, 
arXiv:0812.2022 
\bibitem[De Propris {et~al.}(2002)]{DePropris2002} De Propris, R., {et~al.}\ 2002, \mnras, 329, 87  
\bibitem[Dressler(1980)]{Dressler1980a} Dressler, A.\ 1980, \apjs, 42, 565  
\bibitem[Dressler {et~al.}(1997)]{Dressler1997} Dressler, A., {et~al.}\ 
1997, \apj, 490, 577 
\bibitem[{Ebeling {et~al.}(1996)}]{Ebeling1996} Ebeling, H., Voges, W., Bohringer, H., {et~al.} 1996, MNRAS, 281, 799  
\bibitem[{Ebeling {et~al.}(1998)}]{Ebeling1998} Ebeling, H., Edge, A.~C., Bohringer, H., {et~al.} 1998, MNRAS, 301, 881  
\bibitem[{Ebeling {et~al.}(2000)}]{Ebeling2000} Ebeling, H., Edge, A.~C., Allen, S.~W., {et~al.} 2000, MNRAS, 318, 333  
\bibitem[Fasano {et~al.}(2000)]{Fasano2000} Fasano, G., Poggianti, B.~M., Couch, W.~J., Bettoni, D., Kj{\ae}rgaard, P. \& Moles, M.\ 2000, \apj, 542, 673  
\bibitem[Fasano {et~al.}(2006)]{Fasano2006} Fasano, G., {et~al.}\ 2006, \aap, 445, 805  (Paper I)  
\bibitem[Franceschini {et~al.}(1998)]{Franceschini1998} Franceschini, A., {et~al.} 1998, \apj, 506, 600  
\bibitem[Goto {et~al.}(2002)]{Goto2002} Goto, T., {et~al.}\ 2002,  \aj, 123, 1807  
\bibitem[Graham \& Driver(2005)]{Graham2005} Graham, A.~W., \& Driver,  S.~P.\ 2005, Publications of the Astronomical Society of Australia, 22, 118   
\bibitem[Gunn \& Gott(1972)]{Gunn1972} Gunn, J.~E.~\& Gott, J.~R.~I.\ 1972, \apj, 176, 1  
\bibitem[Infante(1987)]{Infante1987} Infante, L.\ 1987, \aap, 183, 177
\bibitem[{Katgert {et~al.}(1996)}]{Katgert1996}  Katgert,  P.,  Mazure, A.,  Perea, J. {et~al}, 1996, \aap, 310, 8  
\bibitem[Kron(1980)]{Kron1980} Kron, R.~G.\ 1980, \apjs, 43, 305
\bibitem[Kurk {et~al.}(2004)]{Kurt2004} Kurk, J.~D., Pentericci,  L., Overzier, R.~A., {et~al.} 2004, \aap,  428, 817  
\bibitem[Lubin {et~al.}(2002)]{Lubin2002} Lubin, L.~M., Oke, J.~B.,  \& Postman, M.\ 2002, \aj, 124, 1905   
\bibitem[Miller {et~al.}(2005)]{Miller2005} Miller, C.~J., {et~al.}\  2005, \aj, 130, 968
\bibitem[Oemler(1974)]{Oemler1974} Oemler, A.~J.\ 1974, \apj, 194, 1
\bibitem[O'Hely {et~al.}(1998)]{Ohely1998} O'Hely, E., Couch,  W.~J., Smail, I., Edge, A.~C., \& Zabludoff, A.\ 1998, Publications of the  Astronomical Society of Australia, 15, 273   
\bibitem[Pimbblet {et~al.}(2001)]{Pimbblet2001} Pimbblet, K.~A.,  Smail, I., Edge, A.~C., {et~al.} 2001, \mnras, 327, 588  
\bibitem[Postman {et~al.}(2005)]{Postman2005} Postman, M., {et~al.}\  2005, \apj, 623, 721   
\bibitem[Robin {et~al.}(2003)]{Robin2003} Robin, A.~C., Reyl{\'e},  C., Derri{\`e}re, S., \& Picaud, S.\ 2003, \aap, 409, 523   
\bibitem[{Smith {et~al.}(2004)}]{Smith2004} Smith, R., Hudson, M., Nelan, J., {et~al.} 2004, AJ, 128, 1558  
\bibitem[Steidel {et~al.}(2000)]{Steidel2000} Steidel, C.~C., Adelberger, K.~L., Shapley, A.~E., {et~al.} M.\ 2000, \apj, 532, 170
\bibitem[{Valentinuzzi {et~al.}(2008)}]{Valentinuzzi2008}  Valentinuzzi, T. {et~al.} 2008, in preparation.
\bibitem[van Dokkum {et~al.}(2000)]{vanDokkum2000} van Dokkum, P.~G., Franx, M., Fabricant, D., {et~al.} 2000, \apj, 541, 95  
\bibitem[Zwicky {et~al.}(1963)]{Zwicky1963} Zwicky, F., Herzog, E.,  \& Wild, P.\ 1963, "Catalogue of galaxies and of clusters of galaxies".  Pasadena: California Institute of Technology (CIT)    
\end{thebibliography}
\end{document}